\documentstyle[11pt,aaspp4]{article}

\def\refitem{\relax}
\def\journref#1#2#3#4#5{\refitem #1 #2, #3, #4, #5}
\def\paspref#1#2#3#4{\journref{#1}{#2}{\pasp}{#3}{#4}}
\def\apjref#1#2#3#4{\journref{#1}{#2}{\apj}{#3}{#4}}
\def\apjsref#1#2#3#4{\journref{#1}{#2}{\apjs}{#3}{#4}}
\def\ajref#1#2#3#4{\journref{#1}{#2}{\aj}{#3}{#4}}
\def\mnrasref#1#2#3#4{\journref{#1}{#2}{\mnras}{#3}{#4}}

\def\kms{km~s$^{-1}$}
\def\etal{{\it et~al.}}

\def\bra{\mathopen{<}} \def\ket{\mathclose{>}}
\let\gta=\gtrsim

\def\mytabcaption#1{\refstepcounter{table}{#1}\\}

\lefthead{Metzger \& Schechter}
\righthead{Galactic Cepheids}

\begin{document}

\title{A Search for Distant Galactic Cepheids Toward $\ell =
60^\circ$\altaffilmark{1}}

\author{Mark R. Metzger\altaffilmark{2,3} and Paul L. Schechter}
\affil{Physics Department, Room 6-216, Massachusetts Institute of
Technology, \\
    Cambridge, MA 02139 \\
 {\tt mrm@grus.caltech.edu, schech@achernar.mit.edu} }


\altaffiltext{1}{Based in part on observations using the 1.3m McGraw-Hill
telescope at the Michigan-Dartmouth-MIT Observatory}
\altaffiltext{2}{Present Address:  California Institute of Technology,
Astronomy 105-24, Pasadena, CA 91125} 
\altaffiltext{3}{Visiting Astronomer, Kitt Peak National Observatory. 
KPNO is operated by AURA, Inc.\ under contract to the National Science
Foundation.} 


\begin{abstract}

We present results of a survey of a 6-square-degree region near $\ell =
60^\circ$, $b = 0^\circ$ to search for distant Milky Way Cepheids.  Few MW
Cepheids are known at distances $\gta R_0$, limiting large-scale MW disk
models derived from Cepheid kinematics; this work was designed to find a
sample of distant Cepheids for use in such models.  The survey was conducted
in the V and I bands over 8 epochs, to a limiting $I \simeq 18$, with a
total of almost 5 million photometric observations of over 1 million stars.
We present a catalog of 578 high-amplitude variables discovered in this
field.  Cepheid candidates were selected from this catalog on the basis of
variability and color change, and observed again the following season.  We
confirm 10 of these candidates as Cepheids with periods from 4 to 8 days,
most at distances $>3$ kpc.  Many of the Cepheids are heavily reddened by
intervening dust, some with implied extinction $A_V > 10$ mag.  With a
future addition of infrared photometry and radial velocities, these stars
alone can provide a constraint on $R_0$ to 8\%, and in conjunction with
other known Cepheids should provide good estimates of the global disk
potential ellipticity.

\end{abstract}




%

\section{Introduction}

Analysis of the kinematics of classical Cepheids in the disk of the Milky
Way has provided one of the most accurate means of measuring fundamental
parameters of the Galactic disk potential.  Since its first application by
\nocite{joy39} Joy (1939), the catalogs of known Cepheids have grown and
included increasingly higher quality data.  Recent measurements of
fundamental parameters such as the distance to the Galactic center, $R_0$,
and the local circular rotation speed $v_{circ}$, have yielded estimates to
better than 5\% given certain model assumptions (\cite{pmb94}; \cite{mcs97},
hereafter MCS).

The distribution of known Cepheids in the Galactic disk, however, is quite
lopsided: a large fraction of the known distant Cepheids, and hence most of
the leverage in determining $R_0$ from kinematic models, lies in the region
$270^\circ < \ell < 360^\circ$ (in the southern hemisphere).  MCS showed
that by adding only eight well-placed Cepheids to the models, the
uncertainty in the measurement of $R_0$ could be significantly reduced, and
most of the weight in the distance measurement rested on the new Cepheids.
A peculiarity in the properties of the region containing most of the new
Cepheids not taken into account by the models (such as a streaming motion,
unusual dust properties, etc.)  might systematically skew the estimate of
$R_0$.  To confirm the distance measurements and test the validity of the
model assumptions, one would like to obtain additional Cepheids with good
$R_0$ leverage in other regions of the disk.

One source of systematic uncertainty in kinematic measurements of $R_0$ is a
possible large-scale deviation of the true rotation curve from axisymmetry
(e.g.  \cite{bs91}; \cite{kt94}); many traditional kinematic models have
assumed axisymmetry.  MCS found that the existing Cepheid sample is
inadequate for measuring many deviations from axisymmetry, and suggested a
two-pronged strategy to extend the sample so that each of two ellipticity
components could be measured directly.  To address the issue of obtaining
additional Cepheids to help reinforce our measurement of $R_0$, and to
provide additional constraints on the rotation curve ellipticity, we
undertook a survey for Cepheids toward $\ell = 60^\circ, b = 0^\circ$ where
few distant Cepheids were known.  The survey is similar in many respects to
the one conducted in the southern hemisphere by \nocite{cks91} Caldwell,
Keane, and Schechter (1991, hereafter CKS), and the design of this survey is
based in part on lessons learned from the CKS survey.  Wide-area surveys
requiring accurate photometry of a large number of stars have only recently
become possible due to the availability of large charge-coupled device
cameras.  The new CCDs cover a large area of sky while providing enough
spatial resolution to allow accurate photometry, even in the crowded fields
associated with the Galactic plane.

At the time the survey was proposed, one of the best facilities available
was the KPNO 0.9m telescope/Tektronix $2048^2$ CCD combination:  this
configuration can cover a square degree in seven pointings with 0.7
arcsecond sampling.  We had initially started a survey during summer
shutdown at the McGraw-Hill 1.3m telescope at MDM Observatory, but the only
detector then available covered an area of sky 20 times smaller and thus
limited the survey to an insufficient total area.  Even with the high
efficiency of the 0.9m telescope, however, only a limited area of the
Galactic plane can be covered in a single observing run.  We therefore took
some care in the design of the survey to maximize the payoff for
Galactic structure study.

\section{Survey Design}

Cepheids that contribute weight to both a more precise measurement of $R_0$
and global rotation curve shape lie at distances of several kiloparsecs
($\gta 0.5 R_0$).  Given a fixed accuracy in measuring distances to the
tracers, \nocite{sack92} Schechter \etal\ (1992) found that for stars lying
near the solar circle, the uncertainty in $\log R_0$ caused by the intrinsic
velocity dispersion of the tracer ($\sim 10$ \kms\ in the disk for Cepheids)
is minimized in the northern Milky Way toward $\ell \sim 35^\circ$.  Cepheids
that lie on the solar ``circle'' (rotating at the same angular speed) have
the additional advantage of constraining $R_0$ independently of the rotation
speed.  Unfortunately, extinction to these stars due to dust can be quite
large:  they lie at a distance of over $1.7 R_0$, and the line of sight
passes within $0.6 R_0$ of the Galactic center.  Further, measuring an
asymmetry in the rotation curve is much simplified by having tracers
symmetric about the Galactic center.  We therefore chose to conduct the
survey near $\ell = 60^\circ$ where few distant Cepheids are known, reducing
the total extinction and complementing the CKS survey toward $\ell =
300^\circ$. 

Limits on the survey latitude can be set based on the measured distribution
of local Cepheids, which have a scale height of 70 pc (\cite{ks63}).  At a
distance of $R_0$ this corresponds to roughly $0{\rlap{.}}^\circ 5$, thus to
find Cepheids at a distance of $R_0$ we should focus on areas with $|b| <
0{\rlap{.}}^\circ 5$.  Indeed, of the Cepheids discovered in the CKS survey,
all but one were within this latitude range.  One might argue that since the
extinction close to the plane is very high, we should avoid $b = 0^\circ$
and look slightly away, improving the depth of our survey.  Unfortunately
the dust is unavoidable:  the vertical scale height of Cepheids is similar
to that of dust, and so to reach distant Cepheids one must necessarily look
through the dust as well.  If one moves out of the plane, the integrated
dust decreases, but the survey becomes less efficient as the Cepheid density
drops as well.

\begin{figure}
	\vspace{3.5in}
	\includegraphics{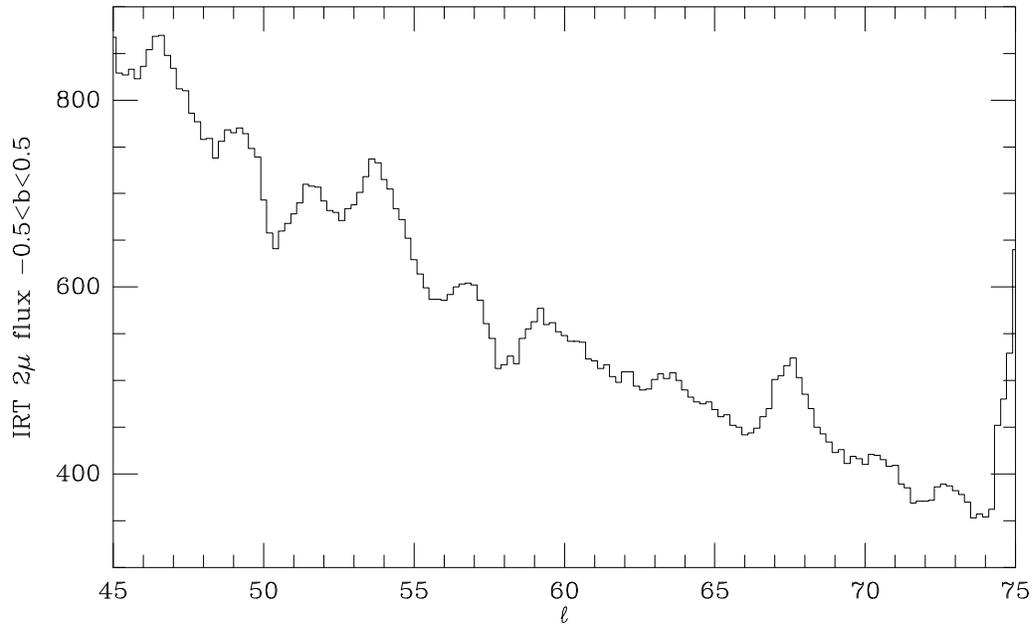}
	\caption[IRT 2$\mu$ Galactic Plane Flux]{
		Spacelab IRT flux in the Galactic plane, integrated over one
		degree in latitude (raw data provided by S. Kent).
		The flux units are arbitrary.  A
		general trend as a function of longitude can be seen
		along with smaller-scale variations.  The effective
		resolution is about 1 degree; some peaks may be due to
		strong unresolved point sources.
		Note the strong peaks near $\ell=68^\circ$ and $\ell =
		54^\circ$, and the low brightness near $\ell = 58^\circ$.
	}\label{fig:irtmap}
\end{figure}

\placefigure{fig:irtmap}

The distribution of dust is not uniform, however, so one can hope to gain an
advantage by choosing lines of sight having relatively low extinction.  CKS
were fortunate to take advantage of one of the least heavily reddened lines
of sight in the inner Galaxy; no comparable region exists near $\ell =
60^\circ$.  We can nonetheless use existing survey data to provide an idea
of which areas have lower extinction, and give these areas priority in our
Cepheid survey.  One method of estimating extinction is to compare the faint
source counts (or total source flux) in different regions.  Since the
stellar luminosity function $\Phi(M)$ is shallower than an $n = 3/5$ power
law, the number of faint sources will decrease dramatically as the
extinction increases.  Surveys at optical wavelengths (e.g. the Palomar
Observatory Sky Survey), however, provide little information on dust at a
distance, as the counts are dominated by stars closer than the Cepheids we
seek, particularly if the total extinction is large.

Near-infrared surface brightness maps are more useful in this regard.  The
extinction is significantly reduced at these wavelengths, and thus the
surface brightness will have a greater contribution from stars at large
distances.  The surface brightness variations (after subtracting a smooth
Galactic component, which varies with longitude) are thus more closely
correlated to the extinction out to many kiloparsecs.  To help select
regions of interest, we examined data from the Spacelab IRT 2$\mu$m survey
of the Galactic plane (\cite{ken92}), which has an effective resolution of
about 1 degree.  Figure~\ref{fig:irtmap} shows a plot of flux integrated
over $|b|<0{\rlap{.}}^\circ5$ as a function of longitude.  Most of the
structure appears on scales larger than the effective resolution, relieving
some concern about contamination from bright point sources.

\begin{figure}
	\vspace{3.5in}
	\includegraphics{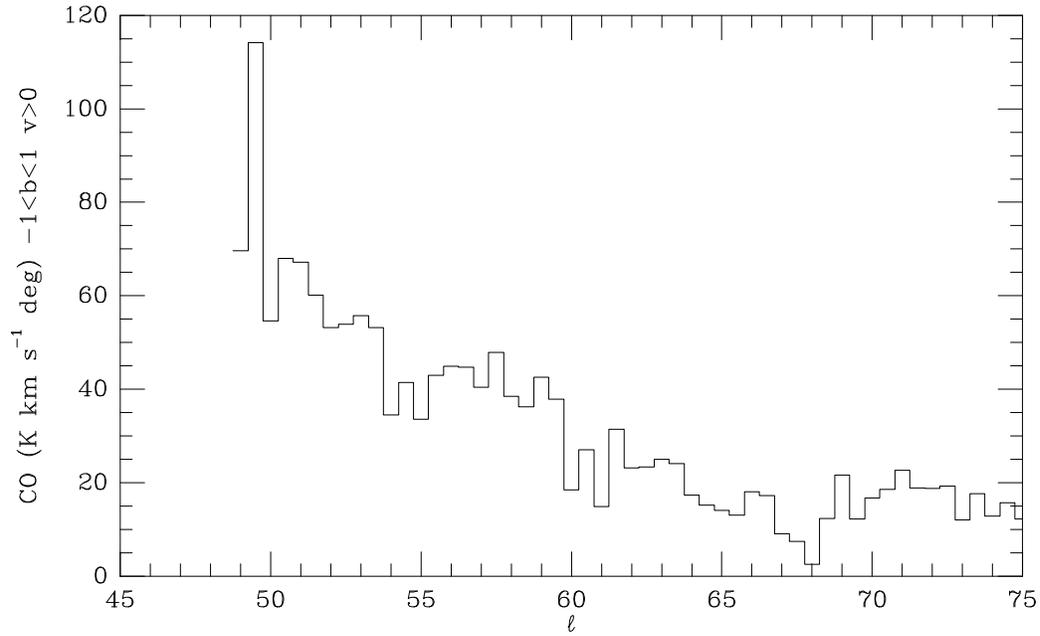}
	\caption[Galactic Plane CO Map]{
		Galactic CO emission integrated over $|b| < 1^\circ$,
		including only gas with positive velocities with respect to
		the LSR.
		Note the similarity of the features between this map and the
		2$\mu$ map, inverted, so that high 2$\mu$ corresponds to low CO
		emission as is expected if the features are caused by
		differential extinction.
	}\label{fig:comap}
\end{figure}

Regions of potentially low extinction can also be mapped using data on
molecular CO emission.  The distributions of gas and dust in the Galaxy have
been shown to be fairly well correlated (e.g.  \cite{hhb76};
\cite{bh78}, \cite{hks81}); CO is a
particularly good tracer of dust as both tend to survive under similar
physical conditions.  We examined data from the survey of \nocite{dam87} Dame \etal\ (1987)
to generate column densities of CO gas as a function of longitude in a
2-degree-wide band at the plane, shown in Figure~\ref{fig:comap}.  A further
advantage of using gas is that the surveys provide column densities
in narrow bands of velocity (1.3 \kms\ in the Dame \etal\
survey), allowing us to select the depth to which we measure the density.
Since all of the gas on the near side of the solar circle has positive
rotational velocity with respect to the LSR, and more distant gas has
negative velocity, by integrating only gas with positive velocity we produce
a total CO column density out to the solar circle.  This provides a better
indication of the total extinction between the Sun and the more interesting
Cepheids.  Several features can be seen in common between the near-infrared
and CO maps: as an example, the strong 2$\mu$ emission near $\ell = 68^\circ$
corresponds to a local minimum of CO column density, precisely what we would
expect if this feature were caused by differential extinction.  

\placefigure{fig:comap}

Using a combination of these two data sets we assigned a relative priority
to different areas along the Galactic plane in the vicinity of $\ell =
60^\circ$, $b = 0^\circ$.  We divided this region of the plane into 98
regions of 1300 arcseconds square with borders aligned north-south. This is
the size and orientation of the Tektronix CCD on the KPNO 0.9m telescope,
allowing for a small overlap between regions (see \S\ref{sec:l60obs}).  A
list of the regions with numeric designations and coordinates is given in
Table~\ref{tab:l60regions}.

\placetable{tab:l60regions}

\subsection{Cepheid Detection}

In addition to selecting areas of the plane with relatively low obscuration,
we can conduct the survey at wavelengths that have lower total extinction.
A longer-wavelength band such as I ($\sim 800$ nm) gives only about 60\% of
the extinction in {\em magnitudes} than suffered at V ($\sim 530$ nm), and
less than half that of B ($\sim 420$ nm) (see, e.g., \cite{ccm89}).  While
the situation improves even more at longer wavelengths, the existing
detectors become significantly smaller:  for surveys in the K-band (2.2
$\mu$m) the largest available detectors had sky dimensions 5 times smaller
(and at poorer resolution) than the large optical CCDs, which would reduce
the survey efficiency by a factor of over 20.  Another competing factor is
the pulsation amplitude, which is significantly larger at blue wavelengths
($\sim$ 1.2 mag at B) than in I ($\sim$ 0.4 mag; \cite{mf91}) or K ($\sim$
0.1 mag; \cite{mcg83}).  Even so, the extra amplitude does not help to find
less heavily reddened Cepheids, where the photometric accuracy in I is more
than sufficient to detect pulsation.  For more heavily-reddened stars the
flux in bluer bands drops dramatically, and requires very long exposures
even to recover the objects.  The best detection sensitivity over a wide
range of distance and extinction is therefore obtained in the reddest bands
we can use.  For optical CCDs this is the I-band, and therefore we followed
CKS in choosing the I band for the primary survey.

The effects of extinction make dynamic range a particularly important issue.
In the absence of extinction, the apparent brightness of a star at $0.1 R_0$
and one at $1.0 R_0$ differ by a factor of 100.  However, it would not be
unreasonable to encounter 5 magnitudes of extinction at I over $0.9 R_0$ (7
kpc) in the inner galaxy, making the distant cousin appear 10,000 times
fainter.  The exposure times were therefore chosen to reach as faint as
possible, while attempting to keep nearby bright Cepheids under saturation.
All but one of the Cepheids discovered in the CKS survey were fainter than
11th magnitude in I, typically with 2 or more magnitudes of extinction.  To
make our bright end cutoff, we tried to insure that we would recover a
10-day period Cepheid at a minimum distance of 0.3$R_0$ under 2 magnitudes
of extinction in I.  Thus we set our exposures to a maximum time that will
place a star of I${}=10.5$ mag at the saturation limit, which was quoted as
240,000 photoelectrons pixel$^{-1}$ for the Tektronix CCD.  Cepheids much
brighter than this would likely have been discovered previously, given the
distribution of known Cepheid magnitudes (\cite{kh88}); the faintest known
Cepheid in our survey region, GX Sge, has $\langle V \rangle = 12.4$ and $I
\simeq 10.3$.

One significant difference between this survey and that of CKS was the
decision to obtain data in the V band for each field at several epochs.
This was motivated by the realization that the characteristic pattern of
color change of a Cepheid over its pulsation cycle is a powerful way to
distinguish Cepheids from other types of variable stars.  In the CKS survey,
stars were selected for follow-up photometry without the benefit of knowing
the color change.  If one were to have this information a priori, many
variable stars could be eliminated before followup photometry was conducted,
and a larger sample of promising candidates could therefore be examined.
However, if V frames are observed throughout the survey, the total area
covered would be cut in half (the V exposures would have to be at least as
long as those in I).  The compromise was to observe V in each field for
every three I observations, providing a reasonable chance of measuring a
color change (which requires at least two points) while reducing sky
coverage by only one quarter.

Another issue is the distribution of individual observations over time.
Identifying a Cepheid requires both detecting its variability at a
sufficient confidence level and recognizing it as a Cepheid from the
properties of its light curve (such as a fast rise/slow decline, color
change, etc.)  While better sampling provides more information on the light
curve shape, CKS showed that 7 epochs of observations were sufficient to
recognize a Cepheid, assuming coverage spaced over the entire cycle.
Cepheids range in period roughly between 3 and 70 days, and the number
distribution is heavily skewed towards shorter periods (Feast \& Walker
1987).  To obtain reasonable phase coverage of longer period Cepheids, the
baseline (number of days between the first and last observation) should be as
long as possible, and observations must be made frequently enough to sample
the short-period Cepheids.  In practice the latter criterion is met without
difficulty, as one can observe each field once per night or on every other
night.  The solution we chose was to observe each field once per night for
four nights, pause for four nights, and observe again for four nights.  This
provides a baseline of 11 days, which should allow detection of Cepheids
with up to 22 day periods (\S\ref{sec:varstars}) while providing adequate
phase coverage for shorter-period Cepheids.  The allocated time was broken
into three nights on, two nights off, and five nights on, which lowered
slightly our sensitivity to long-period Cepheids.

\section{Observations}\label{sec:l60obs}

\placetable{tab:l60olog}

\begin{figure}
	\vspace{7in}
	\includegraphics{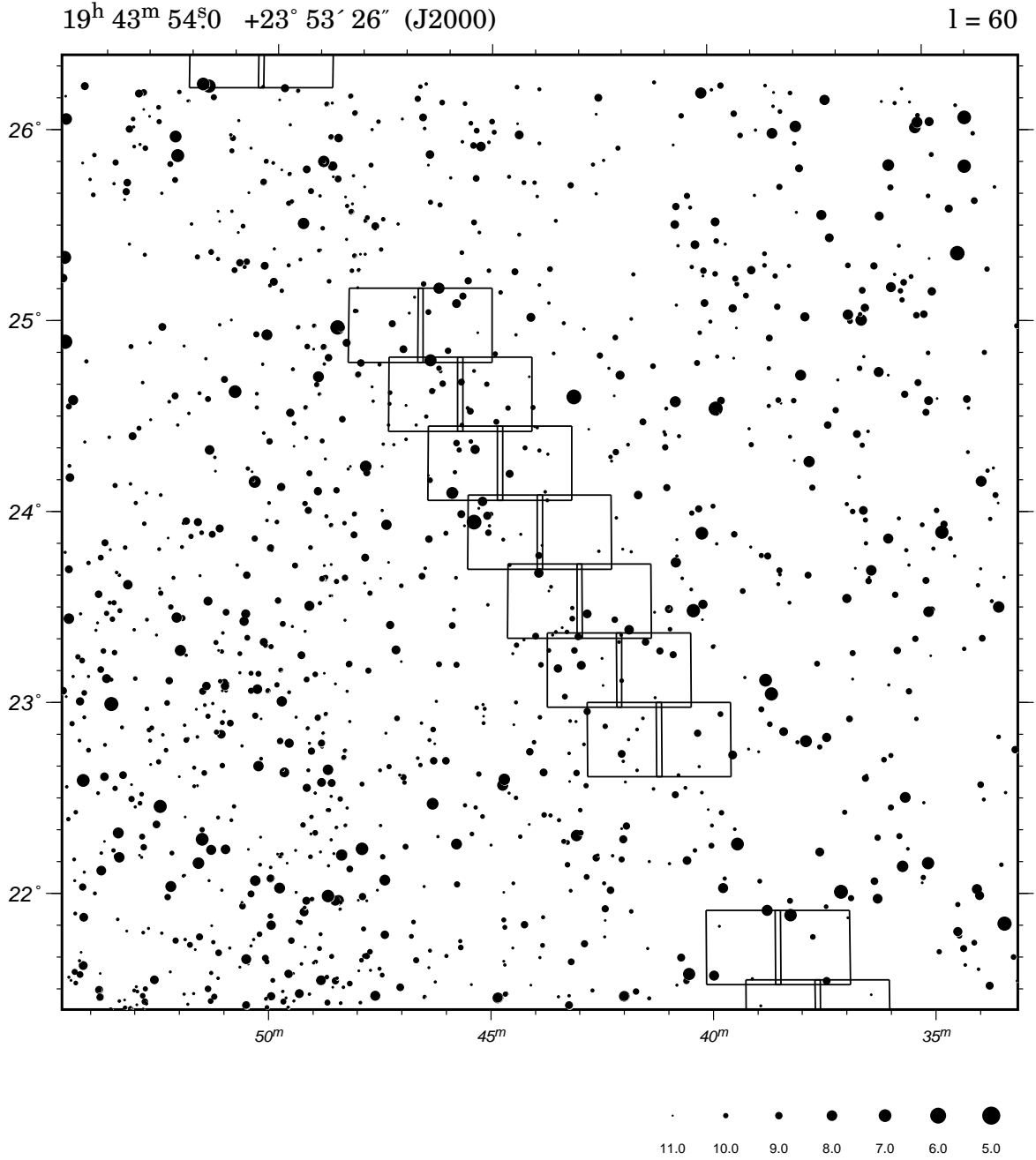}
	\caption[Survey Areas Near $\ell = 60^\circ$]{
		A chart made using data from the HST Guide Star Catalog, showing
		catalog stars brighter than V$=11$ in a
		5 degree square region centered on $\ell = 60^\circ,
		b=0^\circ$.
		Regions
		observed in the Cepheid survey are outlined.  The index in
		the lower right corner shows the point size scale for V
		magnitudes.
	}\label{fig:l60survmap1}
\end{figure}

Observations for the survey were taken with the 0.9m telescope at Kitt Peak
National Observatory on the nights of June 9--11 and 13--18, 1992, using a
Tektronix $2048^2$ CCD.  The detector scale was 0.69 arcseconds per pixel,
giving a field of $\approx 23.5$ arcminutes square (0.15 square degrees).
The regions observed each night and the filters used are listed in
Table~\ref{tab:l60olog}.  We were able to cover a total area of
approximately 6 square degrees over a single night.  The observing
efficiency was limited primarily by the readout time of the chip and the
rate at which the telescope could be moved between fields; a faster readout,
automated repositioning, or possibly scanning along the plane might have
improved observing efficiency.  Figure \ref{fig:l60survmap1} shows a star
map made using the HST Guide Star Catalog (\cite{las90}, hereafter GSC),
with some of the surveyed regions outlined.  Clouds prevented us from
observing for part of night 6 (June 15) and all of night 9 (June 18).  When
telescope hardware problems occurred on two nights that limited the amount
of usable observing time, first priority was given to acquiring the full set
of I observations; for this reason no V observations were taken on night 2.  


\placefigure{fig:l60survmap1}

Exposure times were typically 40 s for I and 60 s for V, but were increased
during periods of poor seeing to compensate for the effective increase in
noise (and due to the reduced danger of saturating bright stars).  The V
filter used was a glass filter from the Kitt Peak ``Harris'' set, the I was
an interference filter.  Traces of the filter response curves are given by
\nocite{sch91} Schoening \etal\ (1991).  Photometric standards of
\nocite{lan92} Landolt (1992) and \nocite{chr85} Christian \etal\ (1985)
were observed at the beginning and end of each night when possible.  Images
of both the twilight sky and an illuminated dome spot were taken each night
in both filters to allow correction of the detector response to an even
illumination level.

\begin{figure}
	\vspace{3.2in}
	\includegraphics{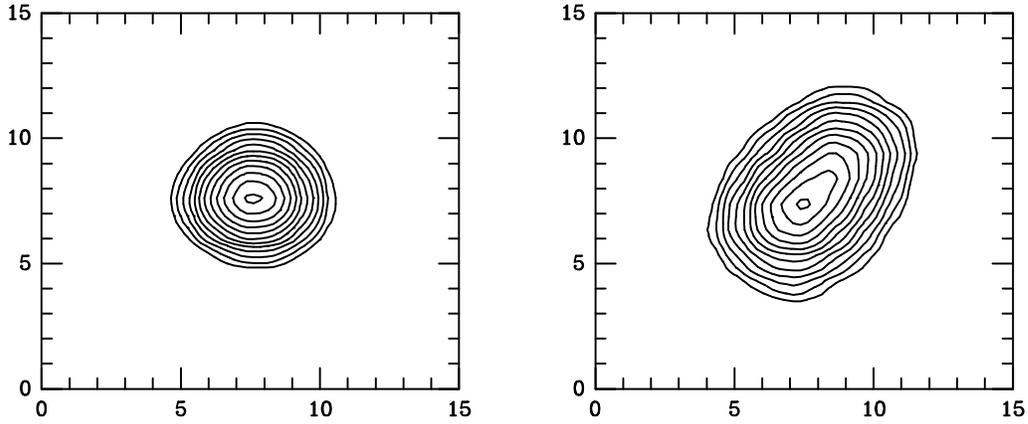}
	\caption[Stellar Point Spread Function]{
		Stellar point-spread intensity functions at the center (left panel)
		and corner (right panel) of the CCD in a single exposure.
		The inside contour is
		chosen near the peak, and subsequent contours are spaced
		logarithmically by factors of $\sqrt{2}$.  The image was
		taken in a period of relatively good seeing
		(1{\rlap{.}}"5).  Axes are shown in units of pixels; the
		boxy appearance is an artifact of the reconstruction.
	}\label{fig:psfthirsix}
\end{figure}

\placefigure{fig:psfthirsix}

A significant problem with the images was the variation of the point spread
function across the chip.  The telescope focal plane was not flat with
respect to the CCD, which caused the focus to vary from the center to the
edge.  The astigmatism (and, as apparent from the images, some coma) present
in the optics produced out-of-focus images that were elongated in the NW-SE
direction on one side of focus, and NE-SW on the other side.  If the focus
was properly adjusted at the center of the chip, the images at the corners
were significantly distorted.  Figure~\ref{fig:psfthirsix} shows the point
spread function near the center and corner of the chip.  Each plot is a
composite of 10 stars, created by subpixelizing, interpolating, centroiding
on the peak, and co-adding a region around each star.  The contours shown
are logarithmic at $\sqrt{2}$ intervals; the third contour from the center
is the half-maximum.  The image distortions are particularly troublesome as
they constantly change:  as the temperature varies through the night, the
focus drifts and has to be re-adjusted.  Between corrections, however, the
PSF will shift shape as the focus shifts, with most areas typically becoming
more elongated.  Even if the focus could be tracked perfectly, the relative
contribution of the astigmatism to the PSF shape is a function of the
atmospheric seeing, which also varies throughout the run.  

The PSF variation across the chip required some extra care in the data
reduction, as described below.  During the observing run, we attempted to
reduce the problem slightly by setting the focus at a compromise position,
where images in a ring around the center were in focus, the center slightly
outside focus, and the corners somewhat inside focus.  The field curvature
problem has since been remedied at the 0.9m telescope with the installation
of a corrector lens in the summer of 1993.

\section{Data Reduction}\label{sec:survreduction}

The bias from each image was computed from a serial overclock region and
subtracted, then the images were corrected for variations in sensitivity
using composite twilight flats; separate composite flats were constructed
for each night.  Some systematic variation was evident between the
individual flats, but it remained $<0.5$\% throughout.  We compared the use
of dome vs. twilight flats to correct the data, and found that the CCD
illumination was quite different between the two.  By comparing the flats to
data images with high night sky levels or images of a globular cluster, M92,
taken at many different positions on the chip, we found the twilight flats
corrected the detector response quite well.  We concluded that the dome
flats suffered from an uneven illumination problem, and were therefore
discarded.

Stars were identified and measured in each of the images using a modified
version of the photometry program DoPHOT (\cite{sms93}).
DoPHOT fits each star to an intensity profile of the form
\begin{equation}
I(x,y) = I_0 {\left( 1 + z^2 + \frac{\beta_4}{2}z^4 + \frac{\beta_6}{2}z^6
	\right)}^{-1},
\end{equation}
\begin{equation}
z^2 = \frac{1}{2} \left( \frac{x^2}{\sigma_x^2} + 2\sigma_{xy}xy +
\frac{y^2}{\sigma_y^2} \right) ;
\end{equation}
where the shape parameters $\beta_4$ and $\beta_6$ are held fixed,
and the other shape parameters $\sigma_x$, $\sigma_y$, and $\sigma_{xy}$ are
allowed to vary when fitting the profile to individual bright stars.
The standard DoPHOT algorithm computes an average shape for the stars in the
image using the means of $\sigma_x$, $\sigma_y$, and $\sigma_{xy}$.  This
average shape is used in fitting each star to measure the flux, and
typically provides a better flux estimate than if the shape parameters were
allowed to vary independently.

This algorithm assumes that the PSF is constant; if the PSF varies across
the chip, the average PSF will not fit any of the stars well, introducing
additional photometric error.  Worse, the PSF at the edge of the chip shown
in Figure~\ref{fig:psfthirsix} is so elongated that it fits better to {\em
two} average PSFs than one, and DoPHOT will happily split every star in the
corner into two components.  We therefore modified the standard DoPHOT
algorithm, to allow the average shape of the point spread function to vary
as a function of position on the chip.  Rather than taking a straight
average for PSF shape parameters, we fit a second order two-dimensional
polynomial for each shape parameter as a function of position.  The flux for
each star was obtained by fitting a PSF using the shape parameters
$\sigma_x(x_0,y_0)$, $\sigma_y(x_0,y_0)$, and $\sigma_{xy}(x_0,y_0)$
obtained from the 3 independent fit functions.  The parameters $\beta_4$ and
$\beta_6$ were fixed at 1.0 and 0.5, respectively.

The second-order polynomial fit for the shape parameters turned out to be
insufficient to match the PSF variation across the entire chip.  A better
match was obtained by breaking up a single 2048$^2$ image into five $1124^2$
tiles (four quadrants plus an overlapping center), fitting the roughly
monotonic PSF variation in each quadrant separately.  After fitting for
positions and fluxes for each star, the catalogs of the individual quadrant
``tiles'' were combined into a single catalog.  The center tile was used as
a reference, and each corner quadrant's overlapping stars were matched to
the reference.  A mean magnitude offset computed from these stars was
applied to each corner tile, to keep the relative photometric system
commensurate between tiles.  This correction was small, typically 0.005 mag
and not exceeding 0.013 mag.

\subsection{Catalog Construction}

To match stars between different observational epochs, the centroid
positions of $\sim 200$ bright stars per field were computed and
cross-referenced between fields.  A transformation consisting of an (x,y)
offset and a linear 2x2 matrix was computed from the coordinates, and the
transformation was used to map the remaining stars to the reference
template.  The data of night 2 was used as the initial template, as it had
the best average seeing.  Two objects were considered a potential match if a
box 3 pixels ($\sim 2$'') high, centered on the transformed position of the
candidate object, included the reference object.  If there were no other
reference stars in the box, the match was considered good and the offsets in
RA and Dec were recorded.  If there was more than one reference star in the
box, the closest star was considered the match and the object was flagged as
possibly confused (``type b'' confusion).  If, however, the best match
reference object for a candidate was also the best match for another
candidate, both candidates are labeled as confused (``type a'' confusion),
and the candidate closest to the reference object is considered the match.
If a candidate object has no match on the reference frame, it is added to the
reference catalog for subsequent use.

After the initial matches were made, a complete reference catalog was produced
using the mean position for each object, and the matching procedure was
repeated.  The scatter in stellar positions between fields taken on
different nights was typically 0.15 pixels rms, or $\sim 0.1$ arcseconds, in
each coordinate.  The number of confused objects was a strong function of the
field crowding, as expected:  the least crowded fields, with $\sim 12,000$
identified objects, typically had 20 confused objects; the most crowded
fields had over 40,000 objects with roughly 300 confused.

In a survey searching for variability it is crucial to ensure that the
observations at different epochs are on the same {\em relative} photometric
system.  We therefore used stars in the field to determine a mean relative
magnitude offset at each epoch.  Since the skies on night 2 were closest to
being photometric (as judged from the standard field analysis), all epochs
were transformed to the night 2 system.  Of the stars matched to the
reference frame, the brightest 5\% (to avoid non-linearity) and faintest
20\% (large scatter) of the stars were eliminated, and the rest used to
derive a mean magnitude offset.  Stars deviating form this mean by more than
5$\sigma$ were eliminated (such stars are likely variable), and a final mean
offset was computed and applied to the field stars to bring them onto the
reference photometric system.  

It was found, however, out that a simple average was not sufficient to bring
the two frames into good relative calibration, as the variation of the PSF
produced errors $\geq 0.05$ mag across the chip.  Most of this effect was
apparently cause by our use of a single offset to transform DoPHOT fit
magnitudes to an aperture system, which is not strictly valid if the PSF
shape varies across the chip.  Another contribution comes from a systematic
difference between the aperture magnitudes for the distorted and normal
PSF.  To allow for this, we fitted a second order two-dimensional polynomial
to the (aperture -- fit) magnitudes as a function of position for night 2,
and used this to correct the magnitudes to full aperture.  The relative
calibration between night 2 and other nights was likewise computed from a
two-dimensional polynomial.  The data was thereby brought to a consistent
system that could be directly calibrated to standard magnitudes.  The
effectiveness of the procedure can be seen both from the formal error of the
offset, $\simeq 0.005$ magnitude, and by measuring $\chi^2_\nu
\simeq 1.0$ for the bright stars (the bright stars have small formal errors
in instrumental magnitude, and thus can be a sensitive test for calibration
errors).  The locations of variable stars are also close to uniform, as
shown below in Figure~\ref{fig:varipos}, though from the excess number of
variables it is evident that the applied correction was insufficient in one
corner.

\begin{figure}
	\vspace{3.75in}
	\includegraphics{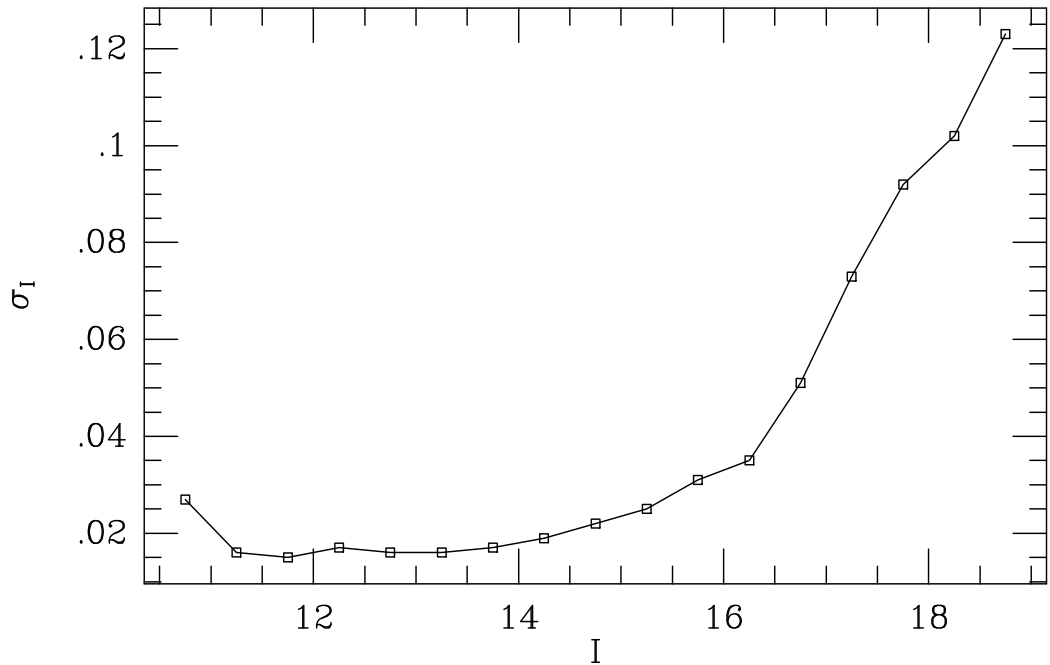}
	\caption[Survey Photometric Accuracy (I Band)]{ 
		Measured dispersion in magnitudes for survey stars,
		plotted as a function of I magnitude.
	}\label{fig:bmgerrs}
\end{figure}

\placefigure{fig:bmgerrs}

Figure~\ref{fig:bmgerrs} shows the photometric errors as a function of I
magnitude, computed from the scatter of non-variable stars over the course
of the survey.  Note that the error bottoms out near 0.016 mag, which is
likely due to residuals from our polynomial fit calibration; for comparison,
the formal error at $I=11.5$ is 0.013 magnitude.  Also note the rise in
error brighter than $I = 11$--this is reflected in both the statistical
dispersion and the formal errors from DoPHOT, and is caused by charge levels
nearing saturation on the detector in nights of good seeing.  While we had
initially designed the survey to avoid this effect down to $I=10.5$, the
detector nonlinearity extended down to levels lower than the instrument
specifications due to changes in the electronics to improve readout speed
(R. Reed, private communication).

The instrumental I band magnitudes from night 2 were converted to the
standard system of \nocite{lan92} Landolt (1992) using exposures of several
fields at the beginning and end of the night.  Corrections were made for an
absolute offset and an airmass term, but no color corrections were applied.
Since many of the stars in our survey are too faint to be detected in the V
band, and thus have no color information, we chose to keep the magnitudes
homogeneous and forego a color correction.  All magnitudes reported in this
section are therefore on an instrumental magnitude system.  The color
corrections required to convert to a standard system are fairly small, at
least over the color range $0 < (B-V) < 2.0$ (see \S\ref{sec:sfollowup}).

Since we do not have V-band data for each region on all nights, the offset
to the standard magnitude system was computed separately for each night.
Light to moderate cirrus obscured our observations on several nights, and
thus our absolute calibration is less well determined for V than for I.  We
can get an idea of how bad the cloud extinction was by looking at the I-band
data taken shortly before or after a V image.  With the exception of night 6,
when the clouds increased steadily until it was no longer possible to
observe, the total extinction from clouds in I was $< 0.07$ mag at all
times.  The absolute calibrations in V therefore should be accurate to $\sim
0.10$ mag.  The photometry of \S\ref{sec:sfollowup} supports this
assessment:  for the fields that overlap the follow-up area, the absolute
calibrations agree to better than 0.05 mag.  For the purposes of identifying
Cepheids, the V photometry is important primarily to measure the color
change, which is not dependent on an accurate absolute calibration.  More
accurate V magnitudes for the Cepheids were then obtained during followup.

Coordinates in equinox J2000 were determined by matching stars in the survey
regions with stars in the GSC in the same manner as we matched the survey
data.  This produces a coordinate transformation from which we can calculate
RA and Dec from the centroid position in pixel coordinates.  Each region had
between 19 and 182 GSC stars, enough to provide a solution good to the
accuracy of the catalog ($\sim 1$'' quoted error, probably somewhat higher
near GSC plate edges).

We initially tried the approach used by CKS to match survey objects, first
by transforming to sky coordinates and performing the match based upon true
sky distances.  We found, however, that errors in star positions near the
edges of the GSC survey plates introduced spurious errors into the
coordinate transformations.  We therefore chose to match objects in pixel
coordinates; those objects in overlap regions are matched using the same
algorithm for {\em inter}-region matching as used above when re-assembling
subregions {\em intra}-region.

\begin{figure}
	\vspace{7.5in}
	\includegraphics{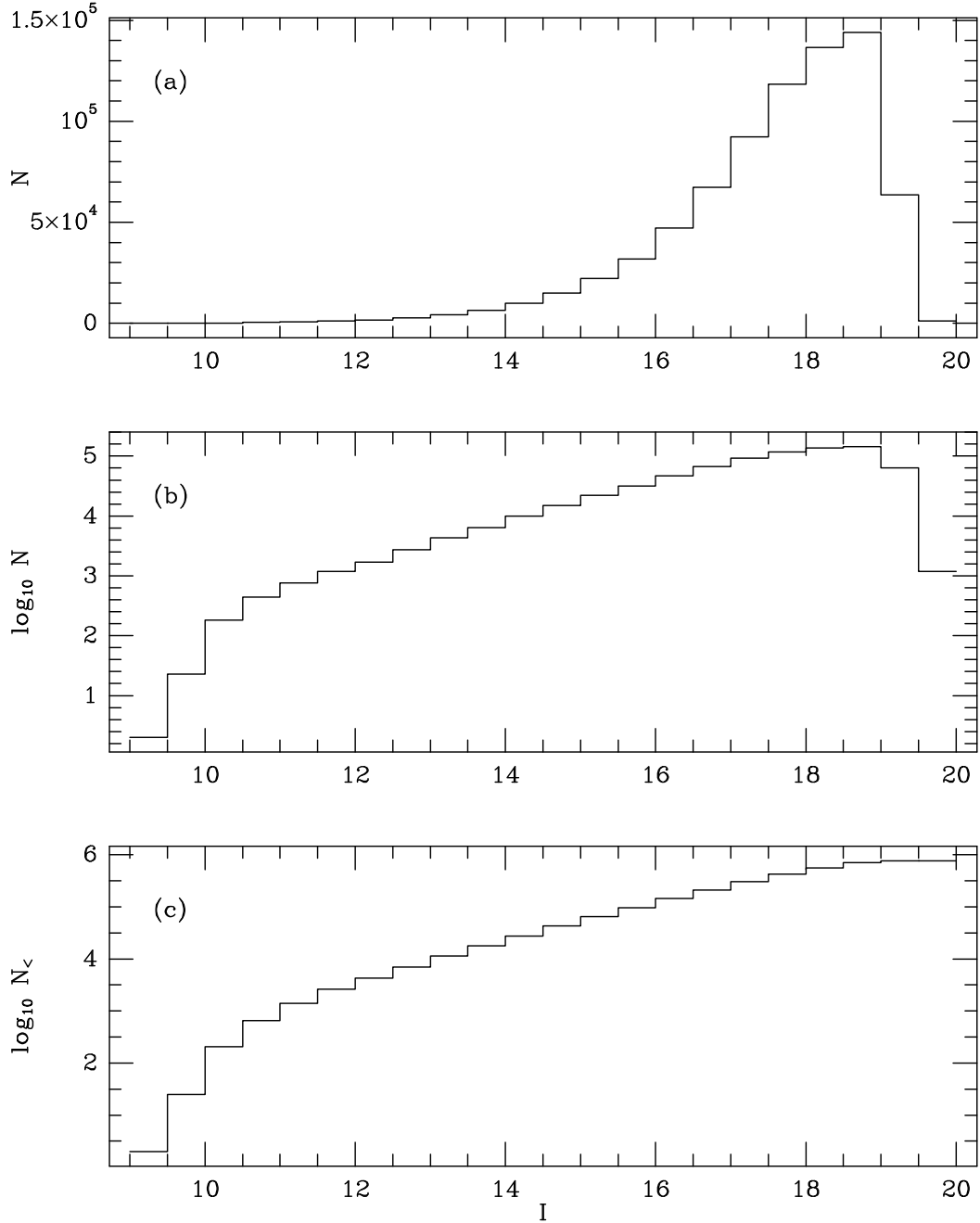}
	\caption[Magnitude Distribution of Survey Stars]{ 
		The magnitude distribution for stars in the survey, counted
		in 0.5 magnitude bins.  (a) The linear count
		distribution as a function of apparent I magnitude. Only
		objects detected on multiple nights are included. (b)
		Same counts as (a) on a logarithmic scale. (c) 
		The cumulative count distribution.  The bright and faint
		magnitude cutoffs are evident; between the two the
		distribution follows roughly a power law in flux,
		though the exponent decreases slightly with fainter
		magnitude.  $d\log N/dI \simeq 0.38$ at $I=14$.
	}\label{fig:magdist}
\end{figure}

\placefigure{fig:magdist}

\subsection{The Catalogs}

The complete survey catalog consists of 4,988,434 photometric measurements
of 1,063,515 stars in an area of roughly 6.1 square degrees.  Of these
stars, 766,816 were detected on three or more nights, and are examined for
variability in \S\ref{sec:varstars}.  Figure~\ref{fig:magdist} shows the
number distribution of catalog stars as a function of apparent magnitude.
The distribution resembles a power law in number vs. flux, and as expected
the slope is somewhat shallower than the dust-free value of $3/5$.  The
completeness begins to fall off for stars of $I \gta 17.5$ and those
brighter than $I=11$, assuming the validity of extrapolating the power law a
small amount at each end.

\begin{figure}
	\vspace{7.5in}
	\includegraphics{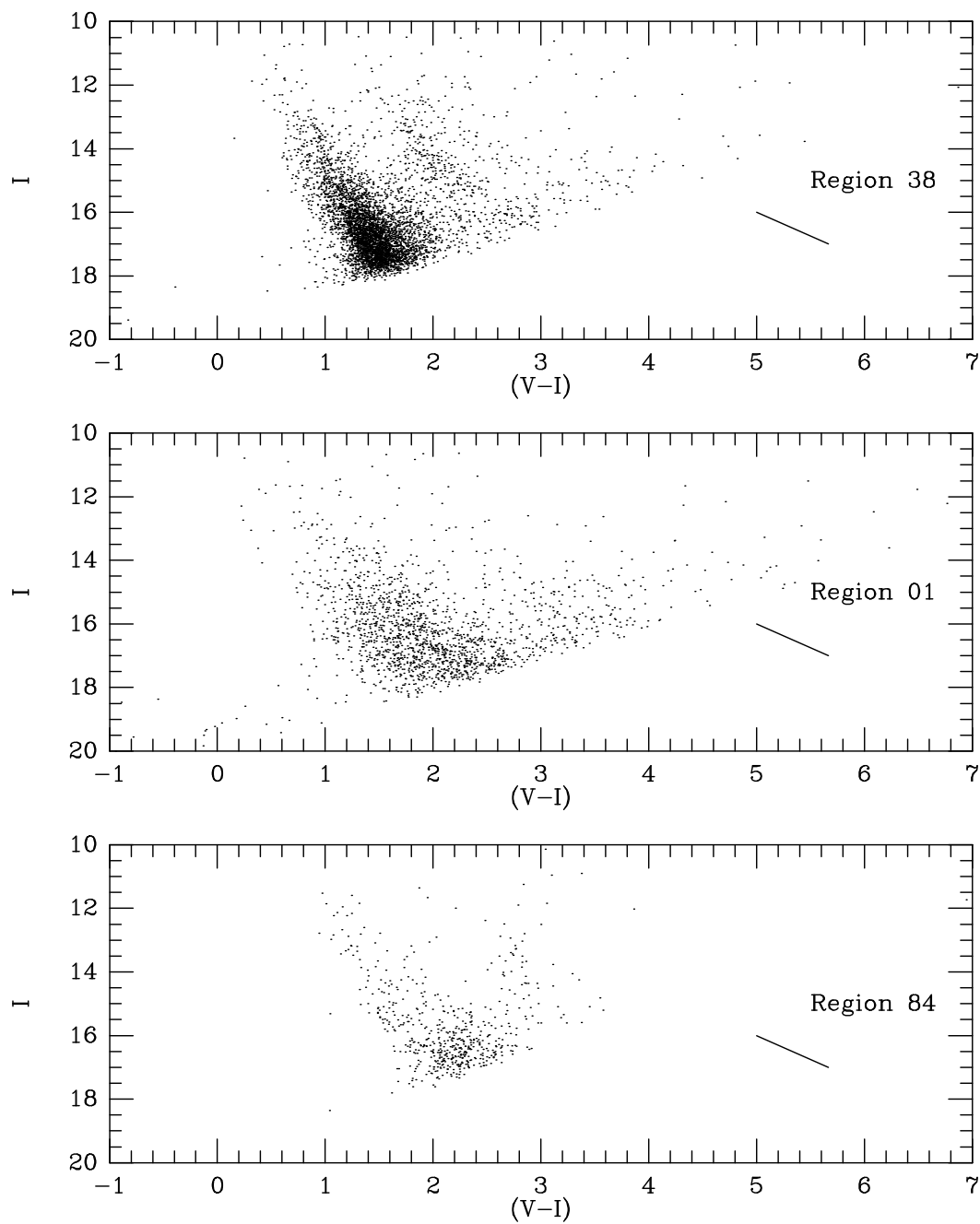}
	\caption[Color-Magnitude Diagrams of Three Survey Regions]{ 
		Color-magnitude diagrams for three survey regions: Region
		38 at $\ell = 67.5^\circ$, Region 1 at $\ell = 60^\circ$,
		and Region 84 at $\ell = 52.9^\circ$.  To reduce crowding,
		one-thrid of the stars with (V--I) colors are plotted for
		each region.  The line at the lower right of each plot shows
		the reddening vector for $A_V = 1$.
	}\label{fig:colorthree}
\end{figure}

\placefigure{fig:colorthree}

Figure~\ref{fig:colorthree} shows color-magnitude diagrams of three survey
regions covering the middle and two extremes of $\ell$ in the survey area.
Not all stars from each region have been plotted, to relieve crowding in the
plot for region 38; the same fraction of stars is shown for each.  Note the
progression to redder colors and fewer identified stars with decreasing
longitude:  only 1,735 stars were identified in both V and I in region 84
($\ell = 53^\circ$), while region 38 ($\ell = 67^\circ$) boasts almost
15,000.  A comparison of the color-magnitude diagrams shows that most of the
difference is due to extinction, which shifts a large number of main
sequence stars below the flux limit.  

\section{Variable Stars and Cepheid Candidates}\label{sec:varstars}

\begin{figure}
	\vspace{3.75in}
	\includegraphics{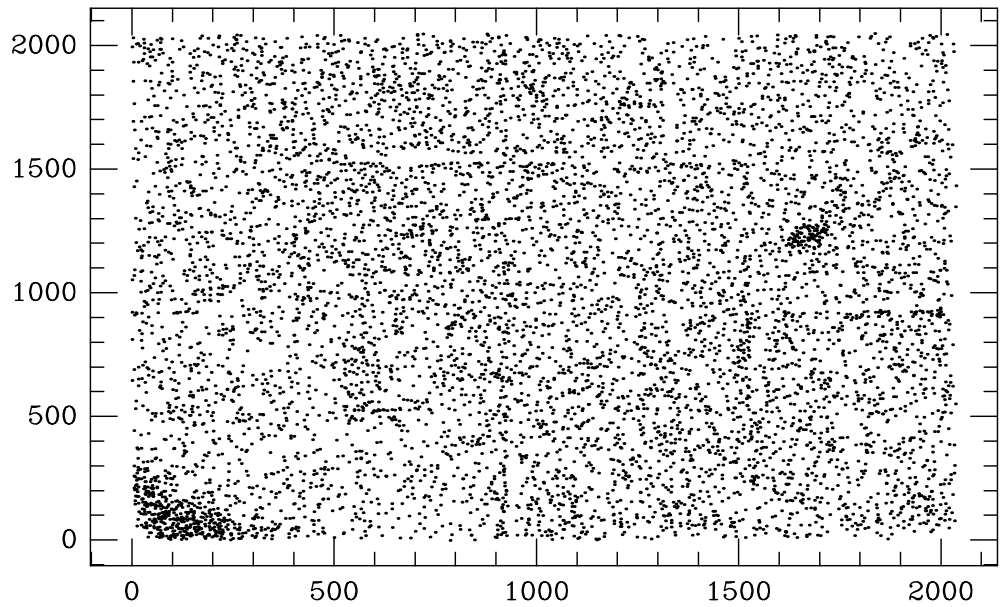}
	\caption[Detector Coordinates of Variables]{ Pixel coordinates on
		the detector for the 7,821 survey stars flagged as
		variable.  Two regions of significant excess can be seen,
		one in the corner where the point spread function was highly
		elongated, and the other near (1700,1200) where the detector
		response may have been nonlinear.  Some signs of the tiling 
		procedure are also evident.
	}\label{fig:varipos}
\end{figure}

\begin{figure}
	\vspace{3.75in}
	\includegraphics{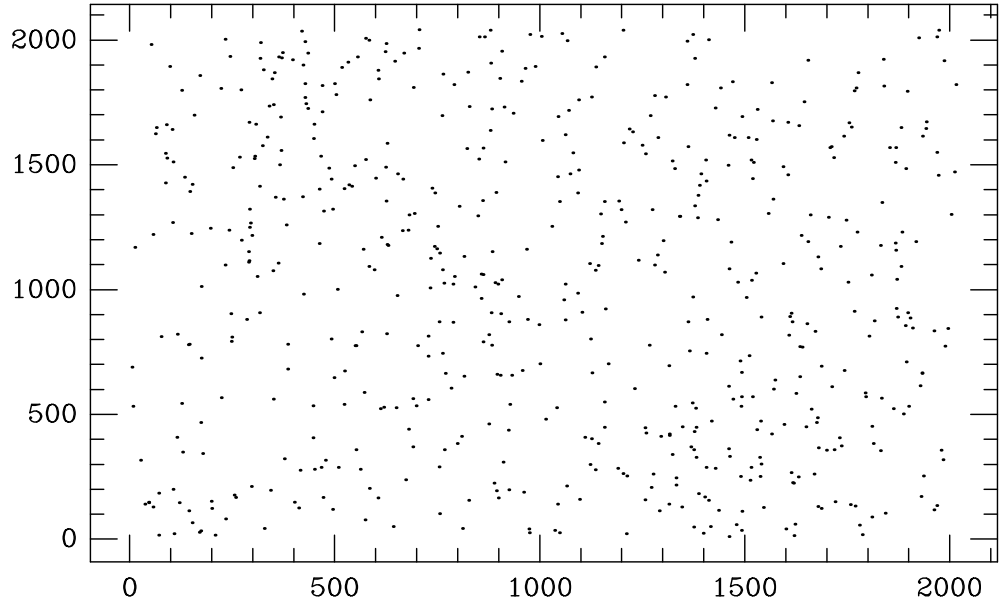}
	\caption[Detector Coordinates of Cepheid Candidates]{
		Detector pixel coordinates of 578 Cepheid candidates.
	}\label{fig:cephpos}
\end{figure}

\placefigure{fig:varipos}
\placefigure{fig:cephpos}

The catalog stars were tested for variability by using the formal errors from
DoPHOT to determine a $\chi^2$ value for each star, under the assumption
that it does not vary.  The criteria for flagging a star as variable were
that it have at least 3 photometric measurements, and that the probability
of exceeding $\chi^2$ if it were not variable, $P(<\chi^2) < 10^{-4}$; this
is similar to the algorithm used by CKS.  Out of the roughly 765,000 stars
in the catalog having three or more measurements, 7,821 were found to be
variable.  Some of these variable stars will be spurious: about 1\% due to the
$\chi^2$ statistics alone.  There are also a higher number of variables
found among survey stars near one corner of the chip, most likely due to the
inability to completely compensate for the systematic offset in photometry.
Figure~\ref{fig:varipos} shows the location on the detector where each
variable star appeared in the survey; in the absence of irregularities, the
distribution should be uniform.  Aside from the corner,
there is a small patch with an anomalously large number of variables.  This
is near a location of low intrinsic response on the detector, which we can
compensate for properly only if the detector maintains a linear
response in that region.  Slightly nonlinear response leads to errors in
the photometric calibration, and could produce the spurious excess of
variables seen.

\placetable{tab:highamp}

Further criteria were placed on the variable star catalog
to extract a subset having 
a sufficiently high variability amplitude to potentially be a Cepheid.  The
typical amplitude of a Cepheid in I is about 0.4 magnitude peak-to-peak; if
we approximate a Cepheid by a continuous sine function, we can calculate the
RMS variability amplitude expected of a Cepheid as follows.  For a full
cycle we have 
\[
\sigma_I^2 = \frac{\int_0^{2\pi} \sin^2 x dx}{\int_0^{2\pi} dx} \frac{A^2}{4} =
0.125 \, A^2 \, ;
\]
for a Cepheid observed over the first half of the cycle we find
\[
\sigma_I^2 = \frac{\int_0^{\pi} {\left( \sin x - \frac{2}{\pi}
\right)}^2 dx}{\int_0^{\pi} dx}
\frac{A^2}{4} = 0.024 \, A^2 \, .
\]
A Cepheid will therefore have $\sigma_I \simeq 0.14$ mag observed over a full cycle and
$0.06$ mag over one half cycle.  We therefore adopted a lower limit of 0.06 mag
for the RMS variability in I, to catch the longest period Cepheids, and a
further criterion of $P(<\chi^2) < 10^{-14}$ to eliminate spurious candidates.
The final selection criterion
is that the star must still qualify as a variable star ($P(<\chi^2) <
10^{-4}$) after any single point in the light curve is removed.  
Strong single-point eclipsers are thus eliminated from further consideration.
The resulting catalog of high-amplitude variables contains 578 stars, and is
presented in Table~\ref{tab:highamp}.

The number of high-amplitude variables based upon the above criteria is too
large to readily acquire additional photometry or spectra for each
candidate.  As a further selection criterion, if a candidate's color change
was $3\sigma$ inconsistent with a $dV/dI$ slope of 1.5, it was excluded.
This removed about 200 of the candidates, though a large fraction of the
candidates do not have color information (see Table~\ref{tab:highamp}) so
this criterion could not be used.  The final selection was made by visually
inspecting light curves for each variable.  Stars showing light curves
inconsistent with a Cepheid's were eliminated, primarily if the amplitude
was too high or the rise time was much slower than the decline.  Light
curves were independently evaluated by the authors in a double-blind
fashion.  In cases where there was disagreement as to the promise of a
candidate, we tried to be inclusive.  In the end, the list of candidates was
cut to 40 of the most promising stars, for which we could perform follow-up
photometry in a single observing run.  Other stars not included may still be
proven to be Cepheids in the future.

Stars selected for follow-up photometry are indicated in
Table~\ref{tab:highamp}.  The ``B'' team candidates---a second set that
looked promising but which we were unable to observe---are also indicated in
the table, and are worthy of further photometry.  Stars that were classified
as high-amplitude variables and showed a monotonic increase or decrease in
brightness are also noted in the table.  Most are likely late-type long
period variables.  Some may be long-period Cepheids, but direct follow-up
photometry was not performed; future spectroscopy could help to
differentiate the Cepheids without requiring multiple epochs.

Most of the variable stars from the General Catalog of Variable Stars, 4th
Edition (\cite{kh88}, hereafter GCVS) in our survey area were recovered,
including GU Vul (W UMa-type), EW Vul (RR Lyr), KN Vul (W UMa), V1023 Cyg
(Algol), and GX Sge ($\delta$ Cep).  A known Cepheid with $\bra I
\ket = 10.2$, GX Sge was at the bright end of our survey limit and was
barely recovered (on two of the nights with the best seeing, it had
saturated the detector).  We found it 1 arcminute away from its reported
position in the GCVS, a much larger offset than the catalog coordinate
precision, but otherwise matching in magnitude and period.  We measure GX
Sge at RA 19h 31m 10.5s, Dec +19$^\circ$ 15' 25" (J2000).  Other GCVS stars
recovered were found at the published locations to within quoted
uncertainties.  The GCVS stars not recovered included V1022 Cyg, a
semi-regular variable with a period of 60 days, which is slightly too bright
at I to be recovered in the survey; GK Vul, a semi-regular with no listed
period, possibly too long for the variability to be detected in this survey;
and CQ Vul, a slow irregular variable.


\subsection{Follow-up Photometry}\label{sec:sfollowup}

Additional photometry of the best Cepheid candidates was obtained on the
nights of May 28--June 7, 1993 at the 1.3 m McGraw-Hill telescope of the
Michigan-Dartmouth-MIT Observatory.  Images were obtained in both V and I
bands using a Tektronix $1024^2$ CCD (``Charlotte''; see \cite{mtl93}
for a general description of the MDM CCD systems).  The CCD and
filters used were similar to the ones used for the main survey:  in both
cases the CCD used was thinned with 24$\mu$ pixels, and the I filter was the
same type of interference filter used a year earlier.  The pixel size was
0{\rlap{.}}"51, which meant that under the best seeing conditions
(0{\rlap{.}}"9) our images were slightly undersampled.  Conditions were
photometric on several nights, allowing us to improve the absolute
calibration for the Cepheids over the original survey data.  Twilight sky
flat-field images were taken each night, and photometric standards of
\nocite{lan92} Landolt (1992) were taken frequently at multiple zenith angles to allow a
correction for atmospheric extinction.  Each frame was corrected for
variations in detector response and throughput using composite twilight
flats, one composite for each night.  From one night to another and over a
range of exposure levels, the corrected response was linear to $<0.3$\%,
except for a 0.6\% difference before and after a thermal cycle of the dewar.

Instrumental magnitudes were measured using apertures 3{\rlap{.}}"5 in
diameter, and corrected to an effective magnitude for a 10" diameter
aperture using isolated bright stars in the images.  Instrumental magnitudes
were measured for the standards in the same manner, and used to determine
atmospheric extinction coefficients and color corrections to a standard
magnitude system.  The transformations are given by
\[
I = m_{1{\rm e}}^i + 23.253(10) - 0.121(20)[\sec z - 1.0] + 0.017(8)[V-I]
\]
and
\[
V = m_{1{\rm e}}^v + 23.683(10) - 0.215(19)[\sec z - 1.0] - 0.014(7)[V-I] \, ,
\]
where $m_{1e}^b = -2.5 \log_{10}(f^b)$, $f^b$ is the corrected 10" flux
in e$^-$ s$^{-1}$.  Colors of standards used to compute these
relations were in the range $-0.21 < (V-I) < 1.76$ and were linear to within
the errors.  Most of our target stars are outside this range, therefore
we have extrapolated this relation to all colors.  Such an extrapolation
is uncertain, however, and should be viewed with caution.  Fortunately the 
color terms are small, and we expect that for $(V-I) < 3.5$ the uncertainty
should be smaller than the typical photometric error for all but the brightest stars.

\subsection{New Cepheids}

\placetable{tab:newcepsumm}

\begin{figure}[p]
	\vspace{7.75in}
	\includegraphics{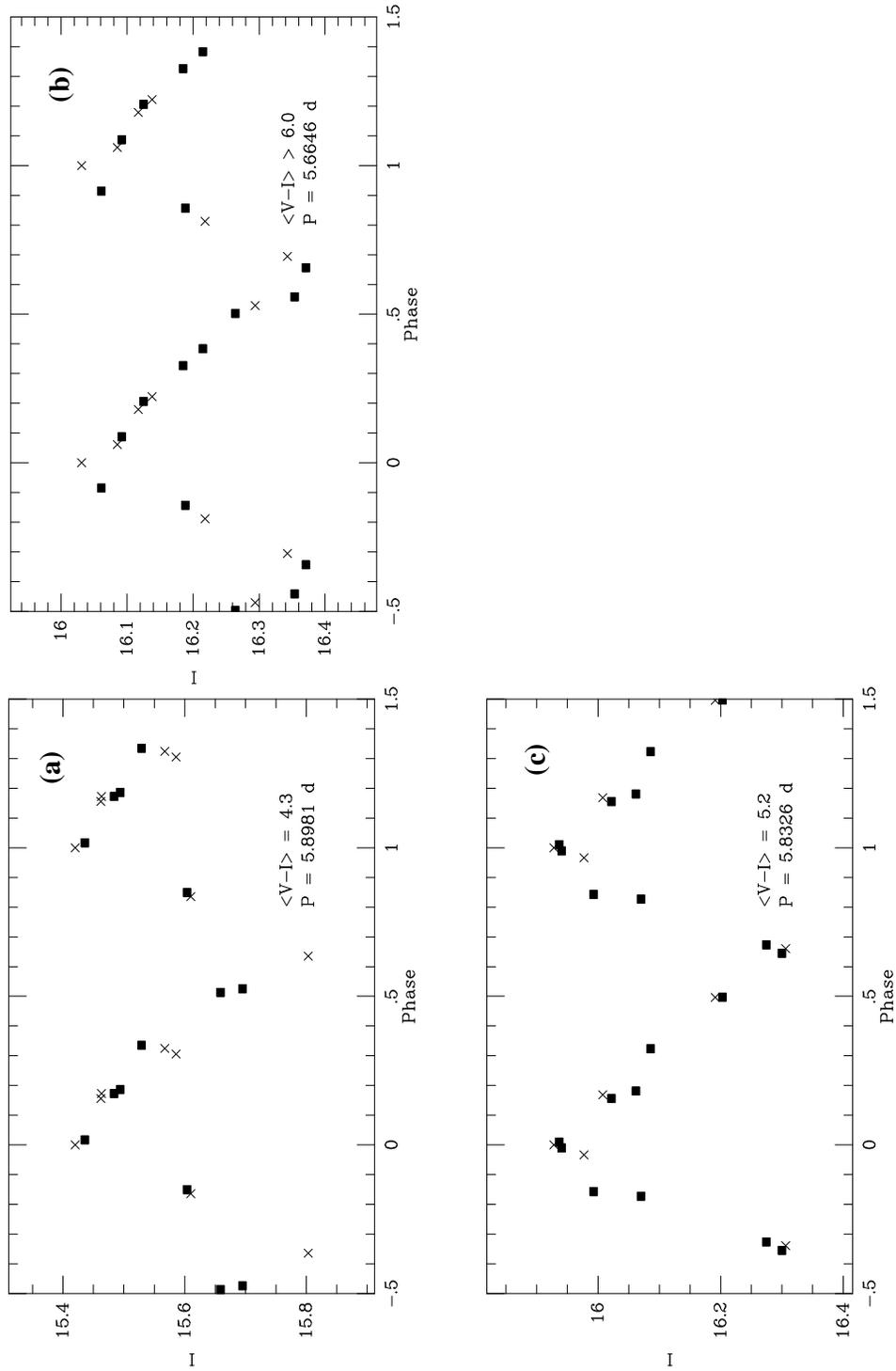}
	\caption[Light Curves: 19313+1901, 19431+2305, 19504+2652]{
		I light curve for the Cepheids (a) 19313+1901; (b)
		19431+2305; and (c) 19504+2652.  Crosses are data
		from the 1992 survey, squares from the 1993 followup.
		These three Cepheids were too faint in the survey data for
		an accurate measurement of the $dV/dI$ slope.
	}\label{fig:newcephlcA}
\end{figure}

\begin{figure}[p]
	\vspace{7.75in}
	\includegraphics{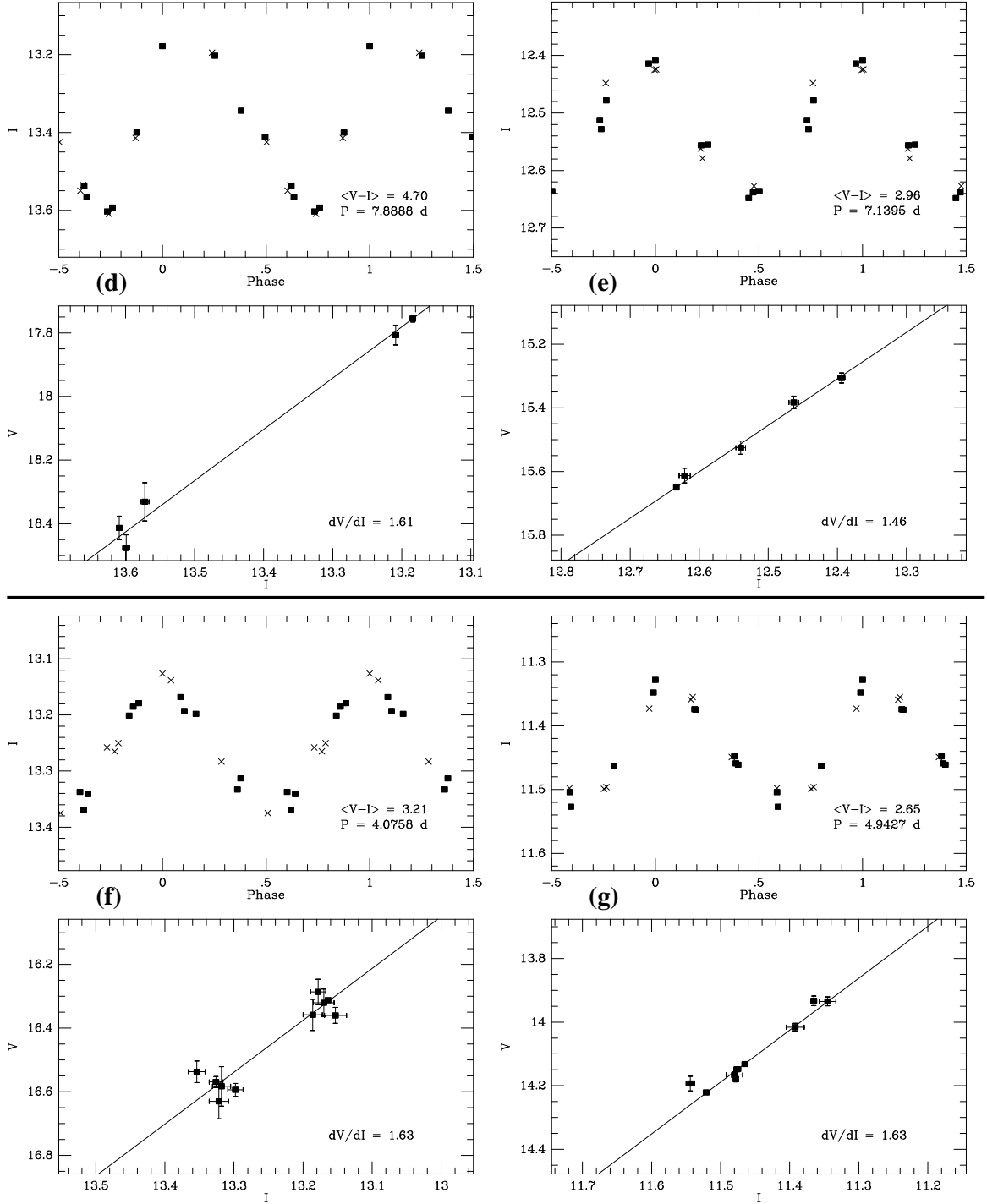}
	\caption[Light Curves: 19430+2326, 19462+2409, 19456+2412, 19468+2447]{
		I light curve and two-band plot for the Cepheids
		(d) 19430+2326; (e) 19462+2409; (f) 19456+2412; and
		(g) 19468+2447.
	}\label{fig:newcephlcB}
\end{figure}

\begin{figure}[p]
	\vspace{7.5in}
	\includegraphics{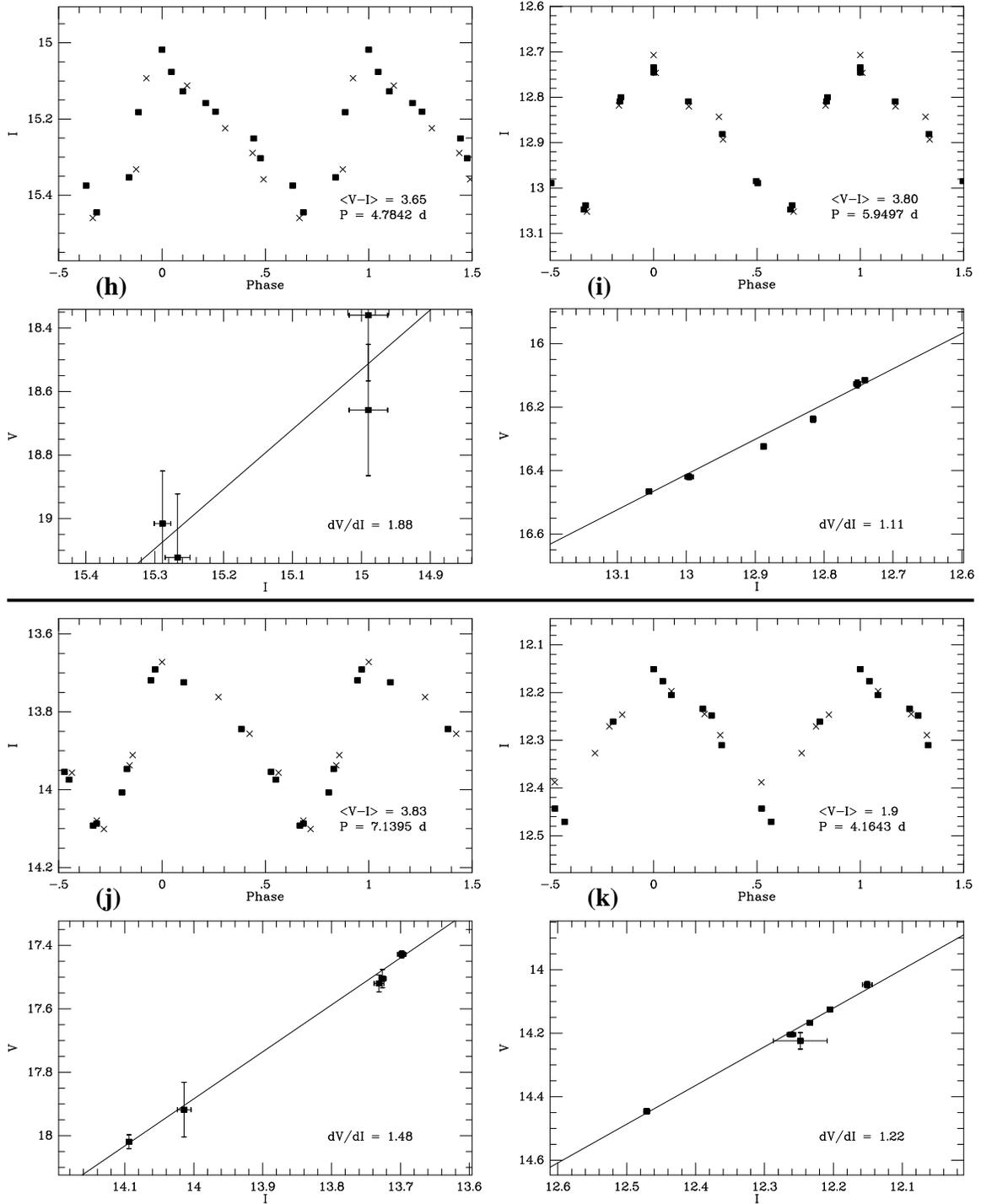}
	\caption[Light Curves: 19462+2501, 19508+2620, 20010+3011, 19286+1733]{
		I light curve and two-band plot for Cepheids 
		(h) 19462+2501; (i) 19508+2620; and (j) 20010+3011.  Panel
		(k) shows the possible Cepheid 19286+1733.
	}\label{fig:newcephlcC}
\end{figure}

Of the 31 stars observed, 10 are confirmed to be Cepheids, with one
additional star likely to be a Cepheid but with an unusual color variation.
Table~\ref{tab:newcepsumm} shows a summary of the data, and Figures
\ref{fig:newcephlcA}--\ref{fig:newcephlcC} show I light curves and V vs. I
color data for the newly-discovered Cepheids. Of the remaining stars, most
had no identifiable periodicity when combined with the original survey data,
down to a period of about 2 days; others appeared periodic but did not have
the appropriate color change for a Cepheid ($dV/dI \gta 1.3$; \cite{mf91},
\cite{avr91}).  The candidate 19450+2400 exhibited clearly
periodic behavior with a long period, and $dV/dI = 1.45$, but the light
curve was too sinusoidal for a Cepheid of that period.

One candidate star (19508+2620) appeared to have a slope of color change too
shallow to be a Cepheid.  Under close examination of a V image taken in good
seeing, however, we discovered a neighboring star close to the candidate
that was roughly equal in brightness to the Cepheid at minimum light.  The V
photometry was contaminated with the light from this star, which caused the
V amplitude to appear shallower and thereby affected the slope of the color
change.  The I light curve is not significantly affected as the Cepheid is
much brighter in I ( due to its relatively red [V--I]$ = 3.8$).  Since this
is an effect of roughly 0.4 magnitudes at V, the overall slope in V if this
star were not present would be $\sim 1.6$, in line with what we expect for a
Cepheid.  Since the I light curve is also clearly consistent with that of a
Cepheid, we are confident that this star is properly classified as such.
Another candidate, 19286+1733, also has a light curve similar to that of a
Cepheid but $dV/dI$ too shallow.  We were unable to identify a contaminating
star in this case, thus we have left its classification as tentative pending
spectroscopic observations.

Periods for the Cepheids were determined using the minimum string length
method (Burke \etal\ 1970, \cite{dwo83}), in a similar way as for southern
hemisphere Cepheids by Avruch (1991).  The observations were folded about a
particular test period, and a string length is computed by summing distances
between points consecutive in phase.  A wide range of test periods were
searched for each star, and the one having the minimum string length is
taken to be the period.  In practice this method is quite sensitive to
photometric errors, and isn't well suited to occasional outliers.  In cases
where this is a problem, however, the minimum string length will usually
correspond to some period that is clearly discordant, and single points can
be deleted and re-fit.  Avruch (1991) performed a Monte Carlo analysis of
period errors associated with this method, but since we have two sets of
observations taken one year apart, our error is dominated by that due to
adding or deleting one full cycle between the two observing seasons.  It is
the one year baseline which gives most of the precision in determining the
period:  a change in period of roughly 1 part in 70 (1 cycle change over a
year for a 5 day period) is not well constrained by data from one year
alone, and thus our periods are accurate only to the 1 part in 70 level.  To
improve the measurement, we would need a third set of observations to remove
the ambiguity, at which point periods should be obtainable to better than 1
part in 1000.

\placefigure{fig:newcephlcA}
\placefigure{fig:newcephlcB}
\placefigure{fig:newcephlcC}

\section{Discussion}

Out of a sample of over 1 million stars in a six square degree area, we have
discovered 10 new Cepheid variables with periods ranging from 4 to 8 days.
As we expected, the extinction toward these stars is significantly higher
than that toward the Cepheids discovered in the CKS survey.  There may also
be a bias against Cepheids outside the 4--8 day period range, which may have
arisen in the qualitative evaluation of the light curves from the survey
(which clearly show the light curve shape for periods in this range).  The
lack of Cepheids with longer periods is likely caused both by being
preferentially excluded due to the short baseline in the survey, and that
they are intrinsically rarer than shorter period Cepheids. 

One of our Cepheids, 19431+2305, is the most heavily reddened Cepheid known
to date.  While we do not yet have an accurate color for this star, we can
barely detect it in a single long V exposure obtained at the 2.4m Hiltner
telescope, which gives puts it at approximately (V--I)$ = 6.3$.  Assuming an
intrinsic color of roughly 0.65 (\cite{mf91}), this implies a
total extinction in V, $A_V$, of roughly 14 magnitudes, using an extinction
law appropriate for Cousins I.  This places the star at a distance modulus
of approximately 12.2, or only 3 kpc.  Such an estimate is only approximate,
as both the calibration of magnitudes to standard bands and the extinction
law are not well determined for stars this heavily reddened.  However,
preliminary observations in the K-band tend to support the 3 kpc distance
estimate.  We have hit a limit on the distance of Cepheids that can be
identified in an I-band survey of the inner disk; going deeper at typical
ground-based seeing results in a significant crowding problem.

\begin{figure}
	\vspace{7.5in}
	\includegraphics{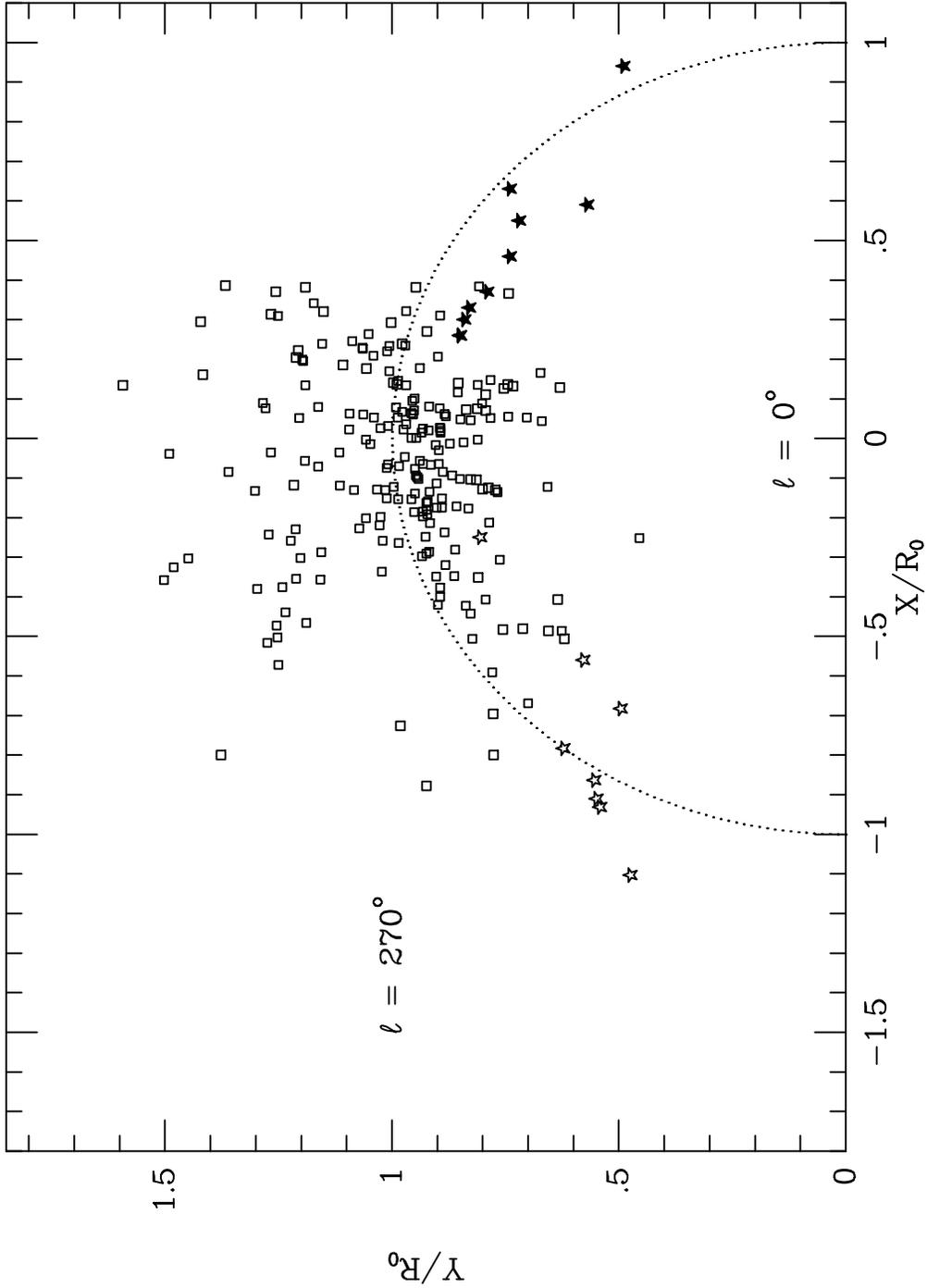}
	\caption[Milky Way Cepheid Positions, New Cepheids]{
		Locations of Cepheids in the Milky Way disk, plotted with
		open squares.  Open stars show Cepheids discovered by CKS in
		a southern hemisphere survey;
		newly discovered northern hemisphere Cepheids from this 
		paper are shown as filled stars.
		Cartesian coordinates are shown in units of $R_0$ with the
                Galactic center at (0,0) and the Sun at (0,1); the solar
		circle ($R=R_0$) is indicated by a dotted line.
	}\label{fig:newcephmap}
\end{figure}

\placefigure{fig:newcephmap}

The approximate positions of the new Cepheids are shown in
Figure~\ref{fig:newcephmap}.  An intrinsic color of (V--I)${} = 0.65$ was
assumed, and an absolute magnitude calibration of $M_I = -3.06(\log P - 1) -
4.87)$ was used to determine distances (\cite{mf91}).  Apparent I magnitudes
of the stars were de-reddened using the relation $A_I = 1.5 E(V-I)$,
following the reddening law of Cohen \etal\ (1981).  Most of the Cepheids
are closer than $R_0$, though we apparently reached the solar circle at
$\ell = 61.2$ and $\ell = 67.4$.  The distances are uncertain primarily due
to the uncertainty in dereddening the apparent magnitudes.  We have also
assumed that each star is a classical (Type I) Cepheid, though with the
available data we are unable to distinguish them from W Vir stars (Type II
Cepheids).  The contamination from W Vir stars should be small, however, as
they are Population II stars and the survey was confined to the disk.

Based on this crude estimate, we see that most of the new Cepheids lie in
regions where none were previously known, and once radial velocities are
measured for these stars, they will provide useful constraints on both $R_0$
and the ellipticity of the rotation curve.  Using the relation given by
Schechter \etal\ 1992,
\[
\frac{dv}{d\ln R_0} = v_o \sin \ell \left[ \frac{d^2 -
d\cos\ell}{\left[ 1+d^2-2d\cos\ell \right] ^{3/2} } \right] \, ,
\]
we find that the most distant Cepheid can by itself provide an estimate of
$R_0$ to
12\%, assuming an intrinsic velocity dispersion of 11 \kms\ in the disk. 
The sample of 10 should yield an $R_0$ measurement with an uncertainty of
8\% once radial velocities and accurate distances are measured.  When
combined with the full sample of known Galactic Cepheids, these stars
can be used to directly test the symmetry of the rotation curve about the
Galactic center to roughly 5\%:  current estimates from the new southern
hemisphere Cepheids are about 5\%, and the measured $R_0$ can be directly
compared in the northern and southern hemispheres.

The high extinction found in the direction of these stars presents two
problems.  The immediate problem is to measure accurate distances to the
newly discovered Cepheids, which can be best accomplished by obtaining
photometry in the near-infrared 2.2$\mu$ K-band.  We have started an
observational program to measure these stars, and have obtained data for
these and many other accessible Cepheids nearby.  The K-band is
significantly less affected by extinction and presents a smaller scatter in
the observed PL relation, both leading to more accurate distances than can
be obtained optically.  Radial velocity measurements are also made more
difficult by the high extinction, but might be obtained from high-resolution
infrared spectra.

It is also evident from the extinction encountered that future surveys for
more distant Cepheids should be conducted in the near infrared.  A survey
concentrating at K, with additional JH photometry to aid identification and
provide reddening estimates, is probably the best strategy for ground-based
surveys, particularly for the inner disk.  At K wavelengths and longer, the
amplitude of variation is roughly 0.3 magnitude, reflecting the change in
surface area (\cite{wel84}).  Even under the equivalent of 30
magnitudes of extinction in V (approximately the extinction to the Galactic
center), a 3-day period Cepheid at a distance of $R_0$ would have an
apparent magnitude of $\sim$13.5 and can easily be measured with the
required photometric accuracy for detecting variability.  Crowding will
become a significant problem, however, and therefore a substantial survey
awaits the availability of large-format infrared arrays (e.g.,
\nocite{hod96} Hodapp et al.  1996) that can simultaneously cover large
areas with sufficient angular resolution. 

\acknowledgments

We thank Hal Halbedel for valuable assistance during our observing at Kitt
Peak, and Bob Barr for working out innumerable last-minute details at MDM.
Thanks also go to Craig Wiegert for his help in acquiring the MDM followup
data.  This work was supported by NSF grants AST-8996139, AST-9015920, and
AST-9215736.

\appendix


\clearpage


\clearpage


\begin{table}
	\centering
	\mytabcaption{Table~\ref{tab:l60regions}. Galactic Plane Regions, $50^\circ < \ell < 70^\circ$}
        \label{tab:l60regions}
	\vspace{2mm}
\newdimen\digitwidth
\newdimen\minuswidth
\newdimen\colonwidth
\setbox0=\hbox{\rm0}
\digitwidth=\wd0
\catcode`?=\active
\def?{\kern\digitwidth}
\setbox0=\hbox{\rm--}
\minuswidth=\wd0
\catcode`!=\active
\def!{\kern\minuswidth}
\setbox0=\hbox{\rm:}
\colonwidth=\wd0
\catcode`<=\active
\def<{\kern\colonwidth}
{ \tabcolsep=0.3cm
\renewcommand{\baselinestretch}{0.87} \small 
\begin{tabular}{ccrcc}
\hline
\hline
\rule{0cm}{10pt}
Region &  \multicolumn{2}{c}{Center} & \multicolumn{2}{c}{J2000} \\
 ID      & $\ell^\circ$ & $b^\circ$?? & RA & Dec \\[0.5mm]
\hline
   01 & 60.09 & -0.16 & 19 44 41.6 & +23 53 26 \\
   02 & 59.91 &  0.16 & 19 43 06.8 & +23 53 26 \\
   03 & 60.51 & -0.16 & 19 45 36.5 & +24 15 06 \\
   04 & 60.33 &  0.16 & 19 44 01.4 & +24 15 06 \\
   05 & 60.92 & -0.16 & 19 46 31.7 & +24 36 45 \\
   06 & 60.74 &  0.16 & 19 44 56.3 & +24 36 45 \\
   07 & 61.34 & -0.16 & 19 47 27.2 & +24 58 23 \\
   08 & 61.16 &  0.16 & 19 45 51.6 & +24 58 23 \\
   09 & 61.76 & -0.16 & 19 48 23.0 & +25 20 00 \\
   10 & 61.58 &  0.16 & 19 46 47.1 & +25 20 00 \\
   11 & 62.18 & -0.16 & 19 49 19.2 & +25 41 36 \\
   12 & 61.99 &  0.16 & 19 47 43.0 & +25 41 36 \\
   13 & 62.59 & -0.16 & 19 50 15.7 & +26 03 10 \\
   14 & 62.41 &  0.16 & 19 48 39.2 & +26 03 10 \\
   15 & 63.01 & -0.16 & 19 51 12.5 & +26 24 42 \\
   16 & 62.83 &  0.16 & 19 49 35.8 & +26 24 42 \\
   17 & 63.43 & -0.16 & 19 52 09.8 & +26 46 14 \\
   18 & 63.25 &  0.16 & 19 50 32.7 & +26 46 14 \\
   19 & 63.84 & -0.16 & 19 53 07.3 & +27 07 44 \\
   20 & 63.66 &  0.16 & 19 51 30.0 & +27 07 44 \\
   21 & 64.26 & -0.16 & 19 54 05.3 & +27 29 12 \\
   22 & 64.08 &  0.16 & 19 52 27.6 & +27 29 12 \\
   23 & 64.68 & -0.16 & 19 55 03.6 & +27 50 39 \\
   24 & 64.50 &  0.16 & 19 53 25.6 & +27 50 39 \\
   25 & 65.09 & -0.16 & 19 56 02.3 & +28 12 04 \\
   26 & 64.91 &  0.16 & 19 54 24.0 & +28 12 04 \\
   27 & 65.51 & -0.16 & 19 57 01.5 & +28 33 28 \\
   28 & 65.33 &  0.16 & 19 55 22.8 & +28 33 28 \\
   29 & 65.93 & -0.16 & 19 58 01.0 & +28 54 50 \\
   30 & 65.75 &  0.16 & 19 56 22.0 & +28 54 50 \\
   31 & 66.34 & -0.16 & 19 59 00.9 & +29 16 11 \\
   32 & 66.16 &  0.16 & 19 57 21.5 & +29 16 11 \\
   33 & 66.76 & -0.16 & 20 00 01.3 & +29 37 30 \\
   34 & 66.58 &  0.16 & 19 58 21.6 & +29 37 30 \\
   35 & 67.18 & -0.16 & 20 01 02.0 & +29 58 47 \\
   36 & 67.00 &  0.16 & 19 59 22.0 & +29 58 47 \\
   37 & 67.60 & -0.16 & 20 02 03.2 & +30 20 02 \\
   38 & 67.42 &  0.16 & 20 00 22.8 & +30 20 02 \\
   39 & 68.01 & -0.16 & 20 03 04.9 & +30 41 16 \\
   40 & 67.83 &  0.16 & 20 01 24.1 & +30 41 16 \\
   41 & 68.43 & -0.16 & 20 04 07.0 & +31 02 28 \\
   42 & 68.25 &  0.16 & 20 02 25.9 & +31 02 28 \\
   43 & 68.85 & -0.16 & 20 05 09.6 & +31 23 38 \\
   44 & 68.67 &  0.16 & 20 03 28.1 & +31 23 38 \\
   45 & 69.26 & -0.16 & 20 06 12.7 & +31 44 46 \\
   46 & 69.08 &  0.16 & 20 04 30.7 & +31 44 46 \\
   47 & 69.68 & -0.16 & 20 07 16.2 & +32 05 52 \\
   48 & 69.50 &  0.16 & 20 05 33.9 & +32 05 52 \\
   49 & 70.10 & -0.16 & 20 08 20.2 & +32 26 56 \\
   50 & 69.92 &  0.16 & 20 06 37.5 & +32 26 56 \\
\hline
\end{tabular}
\normalsize} \\
\vspace{2mm}

\end{table}

\begin{table}
	\centering
	{Table~\ref{tab:l60regions}---{\em Continued}}\\
	\vspace{2mm}
\newdimen\digitwidth
\newdimen\minuswidth
\newdimen\colonwidth
\setbox0=\hbox{\rm0}
\digitwidth=\wd0
\catcode`?=\active
\def?{\kern\digitwidth}
\setbox0=\hbox{\rm--}
\minuswidth=\wd0
\catcode`!=\active
\def!{\kern\minuswidth}
\setbox0=\hbox{\rm:}
\colonwidth=\wd0
\catcode`<=\active
\def<{\kern\colonwidth}
{ \tabcolsep=0.3cm
\renewcommand{\baselinestretch}{0.87} \small 
\begin{tabular}{ccrcc}
\hline
\hline
\rule{0cm}{10pt}
Region &  \multicolumn{2}{c}{Center} & \multicolumn{2}{c}{J2000} \\
 ID      & $\ell^\circ$ & $b^\circ$?? & RA & Dec \\[0.5mm]
\hline
   51 & 59.67 & -0.16 & 19 43 47.0 & +23 31 44 \\
   52 & 59.49 &  0.16 & 19 42 12.5 & +23 31 44 \\
   53 & 59.26 & -0.16 & 19 42 52.8 & +23 10 01 \\
   54 & 59.08 &  0.16 & 19 41 18.5 & +23 10 01 \\
   55 & 58.84 & -0.16 & 19 41 58.8 & +22 48 17 \\
   56 & 58.66 &  0.16 & 19 40 24.8 & +22 48 17 \\
   57 & 58.42 & -0.16 & 19 41 05.1 & +22 26 32 \\
   58 & 58.24 &  0.16 & 19 39 31.4 & +22 26 32 \\
   59 & 58.01 & -0.16 & 19 40 11.7 & +22 04 46 \\
   60 & 57.82 &  0.16 & 19 38 38.2 & +22 04 46 \\
   61 & 57.59 & -0.16 & 19 39 18.6 & +21 42 58 \\
   62 & 57.41 &  0.16 & 19 37 45.3 & +21 42 58 \\
   63 & 57.17 & -0.16 & 19 38 25.7 & +21 21 10 \\
   64 & 56.99 &  0.16 & 19 36 52.7 & +21 21 10 \\
   65 & 56.75 & -0.16 & 19 37 33.2 & +20 59 20 \\
   66 & 56.57 &  0.16 & 19 36 00.3 & +20 59 20 \\
   67 & 56.34 & -0.16 & 19 36 40.8 & +20 37 30 \\
   68 & 56.16 &  0.16 & 19 35 08.2 & +20 37 30 \\
   69 & 55.92 & -0.16 & 19 35 48.7 & +20 15 39 \\
   70 & 55.74 &  0.16 & 19 34 16.4 & +20 15 39 \\
   71 & 55.50 & -0.16 & 19 34 56.9 & +19 53 46 \\
   72 & 55.32 &  0.16 & 19 33 24.7 & +19 53 46 \\
   73 & 55.09 & -0.16 & 19 34 05.3 & +19 31 53 \\
   74 & 54.91 &  0.16 & 19 32 33.4 & +19 31 53 \\
   75 & 54.67 & -0.16 & 19 33 14.0 & +19 09 59 \\
   76 & 54.49 &  0.16 & 19 31 42.2 & +19 09 59 \\
   77 & 54.25 & -0.16 & 19 32 22.8 & +18 48 04 \\
   78 & 54.07 &  0.16 & 19 30 51.3 & +18 48 04 \\
   79 & 53.84 & -0.16 & 19 31 31.9 & +18 26 08 \\
   80 & 53.66 &  0.16 & 19 30 00.6 & +18 26 08 \\
   81 & 53.42 & -0.16 & 19 30 41.2 & +18 04 11 \\
   82 & 53.24 &  0.16 & 19 29 10.1 & +18 04 11 \\
   83 & 53.00 & -0.16 & 19 29 50.7 & +17 42 13 \\
   84 & 52.82 &  0.16 & 19 28 19.8 & +17 42 13 \\
   85 & 52.58 & -0.16 & 19 29 00.5 & +17 20 15 \\
   86 & 52.40 &  0.16 & 19 27 29.7 & +17 20 15 \\
   87 & 52.17 & -0.16 & 19 28 10.4 & +16 58 16 \\
   88 & 51.99 &  0.16 & 19 26 39.8 & +16 58 16 \\
   89 & 51.75 & -0.16 & 19 27 20.5 & +16 36 16 \\
   90 & 51.57 &  0.16 & 19 25 50.1 & +16 36 16 \\
   91 & 51.33 & -0.16 & 19 26 30.8 & +16 14 15 \\
   92 & 51.15 &  0.16 & 19 25 00.6 & +16 14 15 \\
   93 & 50.92 & -0.16 & 19 25 41.4 & +15 52 14 \\
   94 & 50.74 &  0.16 & 19 24 11.3 & +15 52 14 \\
   95 & 50.50 & -0.16 & 19 24 52.0 & +15 30 12 \\
   96 & 50.32 &  0.16 & 19 23 22.1 & +15 30 12 \\
   97 & 50.08 & -0.16 & 19 24 02.9 & +15 08 09 \\
   98 & 49.90 &  0.16 & 19 22 33.1 & +15 08 09 \\
\hline
\hline
\end{tabular}
\normalsize} \\
\vspace{2mm}

\end{table}

\begin{table}
	\centering
	\mytabcaption{Table~\ref{tab:l60olog}. Observation Log}
	\label{tab:l60olog}
	\vspace{2mm}
\newdimen\digitwidth
\newdimen\minuswidth
\newdimen\colonwidth
\setbox0=\hbox{\rm0}
\digitwidth=\wd0
\catcode`?=\active
\def?{\kern\digitwidth}
\setbox0=\hbox{\rm--}
\minuswidth=\wd0
\catcode`!=\active
\def!{\kern\minuswidth}
\setbox0=\hbox{\rm:}
\colonwidth=\wd0
\catcode`<=\active
\def<{\kern\colonwidth}
{ \tabcolsep=0.3cm
\renewcommand{\baselinestretch}{0.92} \small 
\begin{tabular}{c|rrrrrrrr}
\hline
\hline
\rule{0cm}{0.45cm}
 & \multicolumn{8}{c}{Night} \\
 & 1 & 2 & 3 & 4 & 5 & 6 & 7 & 8 \\
\hline
\rule{0cm}{0.45cm}
JD$^a$ & 82 & 83 & 84 & 86 & 87 & 88 & 89 & 90 \\
FWHM $^{\prime\prime}$ & 1.4 & 1.4 & 2.5 & 2.0 & 1.7 & 2.2 & 1.8 & 1.4 \\
\hline
\rule{0cm}{0.45cm}
Region &\multicolumn{8}{c}{Filters Observed}\\[1mm]
01 & VI &  I &  I &  I & VI &    &  I & VI \\
02 &  I &  I &  I &  I & VI &    & VI & VI \\
03 & VI &  I &  I &  I & VI &    &  I & VI \\
04 & VI &  I &  I &  I & VI &    &  I & VI \\
05 & VI &  I &  I &  I & VI &    &  I & VI \\
06 & VI &  I &  I &  I & VI &    &  I & VI \\
07 & VI &  I &  I &  I & VI &    &  I & VI \\
08 & VI &  I &  I &  I & VI &    &  I & VI \\
15 &  I &    & VI &  I & VI &  I &  I & VI \\
16 &  I &  I & VI &  I &  I & VI &  I & VI \\
17 &  I &  I & VI &  I &  I &  I & VI &  I \\
18 &  I &  I & VI &  I &  I &  I & VI & VI \\
19 &  I &  I & VI &  I &  I & VI &  I & VI \\
20 &  I &  I & VI &  I &  I & VI &  I &  I \\
21 &  I &  I & VI &  I &  I & VI &  I &  I \\
22 &  I &  I & VI &  I &  I & VI &  I &  I \\
23 &  I &  I & VI &  I &  I & VI &  I &  I \\
24 &  I &  I & VI &  I &  I & VI &  I &  I \\
35 &  I &  I &  I & VI &  I &  I & VI &  I \\
36 &  I &  I &  I & VI &  I &  I & VI &  I \\
37 &  I &  I &  I & VI &  I &  I & VI &  I \\
38 &  I &  I &  I & VI &  I &  I & VI &  I \\
51 & VI &  I &  I &  I & VI &    &  I & VI \\
52 & VI &  I &  I &  I & VI &    &  I & VI \\
53 & VI &  I &  I &  I & VI &    &  I & VI \\
54 & VI &  I &  I &  I & VI &    &  I & VI \\
55 & VI &  I &  I &  I & VI &    &  I & VI \\
56 & VI &  I &  I &  I & VI &    &  I & VI \\
61 &  I &  I &  I & VI &  I &    & VI &  I \\
62 &  I &  I &  I & VI &  I &    & VI &  I \\
63 &  I &  I &  I & VI &  I &    & VI &  I \\
64 &  I &  I &  I & VI &  I &    & VI &  I \\
65 &  I &  I &  I & VI &  I &    & VI &  I \\
66 &  I &  I &  I & VI &  I &    & VI &  I \\
75 &    &  I &  I &  I &  I &  I & VI &  I \\
76 &    &  I &  I & VI &  I &  I & VI &  I \\
77 &    &    &  I & VI &  I &  I & VI &  I \\
78 &    &    &  I & VI & VI &  I & VI &  I \\
81 &  I &  I &  I & VI &  I & VI &  I & VI \\
82 &  I &  I &  I & VI &  I & VI &  I & VI \\
83 &  I &  I &  I & VI &  I & VI &  I & VI \\
84 & VI &  I &  I & VI &  I & VI &  I & VI \\
\hline
\hline
\end{tabular}
\normalsize} \\  
\vspace{2mm}
$^a$ Julian date minus 2,448,700

\end{table}

\clearpage

\begin{table}[p]
        \centering
        \mytabcaption{Table~\ref{tab:highamp}. High Amplitude Variable Stars}
        \label{tab:highamp}
        \vspace{2mm}
\newdimen\digitwidth
\newdimen\minuswidth
\newdimen\colonwidth
\setbox0=\hbox{\rm0}
\digitwidth=\wd0
\catcode`?=\active
\def?{\kern\digitwidth}
\setbox0=\hbox{\rm--}
\minuswidth=\wd0
\catcode`!=\active
\def!{\kern\minuswidth}
\setbox0=\hbox{\rm:}
\colonwidth=\wd0
\catcode`<=\active
\def<{\kern\colonwidth}
{ \tabcolsep=0.3cm
\renewcommand{\baselinestretch}{0.87} \small 
\begin{tabular}{cccccrcl}
\hline
\hline
\rule{0cm}{10pt}
Catalog ID & \multicolumn{2}{c}{RA (J2000) Dec} & $\langle I \rangle$ &
$\sigma_I$ & $\chi^2_\nu$? & $\langle V \rangle - \langle I \rangle$ &
Notes \\[0.5mm]
\hline
01-00195 & 19 44 48.0 & +23 54 18 & 15.20 & 0.084 &  59.6 &  1.88 & \\
01-00427 & 19 44 24.8 & +23 49 51 & 17.88 & 0.596 &  44.8 &  & \\
01-00637 & 19 44 44.6 & +23 53 35 & 16.20 & 0.206 &  31.2 &  & 1 \\
01-00793 & 19 44 21.4 & +23 47 42 & 16.03 & 0.087 &  28.3 &  1.78 & \\
01-00966 & 19 45 00.3 & +23 56 22 & 16.90 & 0.108 &  15.0 &  2.76 & \\
01-01196 & 19 44 52.2 & +23 52 04 & 16.84 & 0.233 &  24.3 &  2.95 & \\
01-03774 & 19 45 13.3 & +23 50 03 & 13.42 & 0.152 & 185.9 &  1.62 & \\
01-05204 & 19 44 45.2 & +23 44 40 & 17.69 & 0.559 &  34.6 &  & \\
01-07279 & 19 44 33.4 & +23 46 52 & 17.48 & 0.167 &  16.0 &  & \\
01-08549 & 19 44 58.9 & +23 59 47 & 12.40 & 0.123 & 203.7 &  3.35 & 1 \\
01-08687 & 19 45 13.6 & +23 59 35 & 14.09 & 0.076 &  43.1 &  1.44 & \\
01-09100 & 19 45 25.3 & +23 54 50 & 16.11 & 0.061 &  14.7 &  1.99 & \\
01-09318 & 19 45 15.7 & +24 00 51 & 16.42 & 0.080 &  16.5 &  2.19 & \\
01-11655 & 19 44 08.9 & +23 56 58 & 15.07 & 0.099 &  52.7 &  1.69 & \\
01-11809 & 19 43 52.7 & +24 04 34 & 16.07 & 0.124 &  17.6 &  & \\
01-11982 & 19 44 23.6 & +23 59 47 & 17.20 & 0.136 &  12.6 &  & \\
01-12561 & 19 43 49.8 & +23 56 50 & 16.09 & 0.634 & 373.2 &  2.03 & \\
01-19057 & 19 43 59.8 & +24 05 03 & 15.00 & 0.240 &  21.7 &  & \\
02-00285 & 19 43 01.1 & +23 47 31 & 15.40 & 0.149 &  75.9 &  1.96 & \\
02-02092 & 19 42 42.7 & +23 47 38 & 18.48 & 0.553 &  22.9 &  & \\
02-02141 & 19 43 02.3 & +23 45 16 & 13.63 & 0.249 & 667.7 &  1.08 & \\
02-02434 & 19 43 09.5 & +23 47 06 & 16.17 & 0.080 &  17.8 &  2.56 & 2 \\
02-02442 & 19 43 41.1 & +23 46 54 & 16.08 & 0.150 &  47.7 &  2.10 & \\
02-02735 & 19 43 47.6 & +23 54 22 & 17.44 & 0.218 &  19.4 &  2.39 & \\
02-04392 & 19 42 26.5 & +23 44 02 & 16.88 & 0.164 &  25.5 &  2.35 & \\
02-05822 & 19 43 21.4 & +24 02 06 & 15.21 & 0.066 &  32.6 &  1.75 & \\
02-06335 & 19 43 22.3 & +23 59 45 & 16.51 & 0.123 &  22.3 &  2.38 & \\
02-06414 & 19 43 46.0 & +23 54 56 & 16.51 & 0.187 &  64.4 &  2.44 & \\
02-07921 & 19 42 33.4 & +24 01 06 & 15.20 & 0.114 &  11.5 &  1.71 & \\
03-00022 & 19 46 01.7 & +24 14 22 & 12.62 & 0.442 & 2763.6 &  1.12 & \\
03-00055 & 19 45 51.7 & +24 09 11 & 12.96 & 0.094 &  22.8 &  2.81 & \\
03-00092 & 19 45 36.7 & +24 12 09 & 13.21 & 0.087 &  67.6 &  3.20 & 1 \\
03-00360 & 19 45 27.6 & +24 11 39 & 15.55 & 0.175 &  99.1 &  1.71 & \\
03-00433 & 19 45 21.1 & +24 20 06 & 15.43 & 0.100 &  13.5 &  1.65 & \\
03-02298 & 19 45 42.8 & +24 11 36 & 17.50 & 0.300 &  17.6 &  2.21 & \\
03-02725 & 19 45 24.0 & +24 19 34 & 17.88 & 0.752 &  25.8 &  1.16 & \\
03-06544 & 19 46 11.5 & +24 09 04 & 12.51 & 0.090 &  77.1 &  2.72 & 1 \\
03-06552 & 19 46 04.0 & +24 05 36 & 12.06 & 0.252 & 757.2 &  2.36 & \\
03-12117 & 19 45 03.6 & +24 04 16 & 15.25 & 0.147 &  47.0 &  2.70 & 2 \\
03-12186 & 19 44 45.5 & +24 12 58 & 15.64 & 0.090 &  15.9 &  1.58 & \\
03-12425 & 19 45 07.2 & +24 04 07 & 16.11 & 0.082 &  13.5 &  2.85 & \\
03-13446 & 19 45 09.4 & +24 09 24 & 17.94 & 0.201 &  16.3 &  & \\
03-15534 & 19 45 36.9 & +24 26 15 & 13.78 & 0.092 &  15.2 &  & \\
03-15889 & 19 45 55.4 & +24 24 48 & 16.39 & 0.292 &  61.0 &  2.99 & \\
03-15979 & 19 45 40.3 & +24 21 52 & 16.33 & 0.130 &  24.1 &  2.04 & \\
03-15994 & 19 46 15.6 & +24 21 27 & 15.58 & 0.119 &  47.1 &  2.08 & \\
03-17215 & 19 46 11.6 & +24 21 58 & 17.76 & 0.307 &  18.4 &  1.94 & \\
04-00354 & 19 43 49.9 & +24 13 30 & 15.47 & 0.164 & 138.1 &  2.76 & \\
04-01038 & 19 44 17.4 & +24 17 55 & 16.98 & 0.187 &  21.1 &  1.66 & \\
04-01192 & 19 44 02.1 & +24 15 32 & 17.06 & 0.264 &  25.8 &  2.50 & ... \\
\hline
\end{tabular}
\normalsize} \\
\vspace{2mm}

\end{table}

\begin{table}
	\centering
	\mytabcaption{Table~\ref{tab:newcepsumm}. New Cepheids}
        \label{tab:newcepsumm}
	\vspace{2mm}
\newdimen\digitwidth
\newdimen\minuswidth
\newdimen\colonwidth
\setbox0=\hbox{\rm0}
\digitwidth=\wd0
\catcode`?=\active
\def?{\kern\digitwidth}
\setbox0=\hbox{\rm--}
\minuswidth=\wd0
\catcode`!=\active
\def!{\kern\minuswidth}
\setbox0=\hbox{\rm:}
\colonwidth=\wd0
\catcode`<=\active
\def<{\kern\colonwidth}
{ \tabcolsep=0.3cm
\renewcommand{\baselinestretch}{0.87} \small 
\begin{tabular}{ccccccc}
\hline
\hline
\rule{0cm}{10pt}
Star & Catalog ID & \multicolumn{2}{c}{RA (J2000) Dec} & $\langle I \rangle$ &
$\langle V \rangle$--$\langle I \rangle$ &
Period \\
\hline
19313+1901 & 76-13269 & 19 31 15.5 & +19 00 42 & 15.54 & 4.3 & 4.1643 \\
19430+2326 & 52-04808 & 19 42 59.5 & +23 25 35 & 13.37 & 4.7 & 7.8888 \\
19431+2305 & 53-00371 & 19 43 07.3 & +23 04 33 & 16.06 & $>6$  & 5.6646 \\
19456+2412 & 03-00092 & 19 45 36.7 & +24 12 09 & 13.21 & 3.2 & 4.0758 \\
19504+2652 & 18-00380 & 19 50 26.0 & +26 51 44 & 16.09 & 5.2 & 5.8326 \\
19508+2620 & 15-00026 & 19 50 49.3 & +26 19 45 & 12.85 & 3.8 & 5.9497 \\
19462+2409 & 03-06544 & 19 46 11.5 & +24 09 04 & 12.51 & 3.0 & 3.8799 \\
19462+2501 & 08-00258 & 19 46 11.9 & +25 00 33 & 15.28 & 3.7 & 4.7842 \\
19468+2447 & 07-11383 & 19 46 46.9 & +24 46 47 & 11.43 & 2.7 & 4.9427 \\
20010+3011 & 38-09441 & 20 01 01.4 & +30 11 17 & 13.91 & 3.8 & 7.1395 \\
19286+1733 & 84-01800 & 19 28 37.7 & +17 32 36 & 12.24 & 1.9 & 4.1643 \\
\hline
\end{tabular}
\normalsize} \\
\vspace{2mm}

\end{table}

\clearpage

\renewcommand{\bottomfraction}{0.0}
\renewcommand{\floatpagefraction}{0.1}
\setcounter{topnumber}{1}       
\setcounter{bottomnumber}{1}    
\setcounter{totalnumber}{1}     


\clearpage

\clearpage

\end{document}